\documentclass{aastex62}

\graphicspath{{./}{figures/}}

\received{29 October 2019}
\revised{}
\accepted{29 November 2019}
\submitjournal{ApJ}

\shorttitle{Carbon Abundance in the Solar neighbourhood}
\shortauthors{Franchini et al.}

\begin{document}

\title{The Gaia-ESO Survey: Carbon abundance in the Galactic thin and thick disks \footnote{Based on observations collected with the FLAMES instrument at VLT/UT2 telescope (Paranal Observatory, ESO, Chile), for the Gaia--ESO Large Public Spectroscopic Survey (188.B--3002, 193.B--0936).}}

\correspondingauthor{Mariagrazia Franchini}
\email{Mariagrazia.franchini@inaf.it}

\author[0000-0001-5611-2333]{Mariagrazia Franchini}
\affil{INAF - Osservatorio Astronomico di Trieste, Via G. B. Tiepolo 11, Trieste, I--34143, Italy }

\author[0000-0002-3319-6375]{Carlo Morossi}
\affiliation{INAF - Osservatorio Astronomico di Trieste, Via G. B. Tiepolo 11, Trieste, I--34143, Italy }

\author[0000-0003-3168-2289]{Paolo Di Marcantonio}
\affiliation{INAF - Osservatorio Astronomico di Trieste, Via G. B. Tiepolo 11, Trieste, I--34143, Italy }

\author{Miguel Chavez}
\affiliation{Instituto Nacional de Astrof\'isica, \'Optica y Electr\'onica, Luis Enrique Erro 1,  72840 Tonantzintla, Puebla, Mexico}

\author{Vardan Zh. Adibekyan}
\affiliation{Centro de Astrofísica da Universidade do Porto, Rua das Estrelas, 4150-762 Porto, Portugal}

\author{Amelia Bayo}
\affiliation{Instituto de F\'isica y Astronom\'ia, Universidad de Valpara\'iso, Avda. Gran Breta\~na 1111, Valpara\'iso, Chile}
\affiliation{N\'ucleo Milenio de Formaci\'on Planetaria, NPF, Universidad de Valpara\'iso, Chile}

\author{Thomas Bensby}
\affiliation{Lund Observatory, Department of Astronomy and Theoretical Physics, Box 43, SE-221 00 Lund, Sweden}

\author{Angela Bragaglia}
\affiliation{INAF - Osservatorio di Astrofisica e Scienza dello Spazio di Bologna, Via Gobetti 93/3 I-40129 Bologna, Italy}

\author{Francesco Calura}
\affiliation{INAF - Osservatorio di Astrofisica e Scienza dello Spazio di Bologna, Via Gobetti 93/3 I-40129 Bologna, Italy}

\author{Sonia Duffau}
\affiliation{Universidad Andr\'es Bello Departamento de Ciencias Fernandez Concha 700, Las Condes
Santiago Chile}

\author{Anais Gonneau}
\affiliation{Institute of Astronomy University of Cambridge
Madingley Road Cambridge CB3 0HA (UK) }

\author{Ulrike Heiter}   
\affiliation{Observational Astrophysics, Department of Physics and Astronomy, Uppsala University, Box 516, 75120 Uppsala, Sweden}

\author{Georges Kordopatis}
\affiliation{Universit\'e C\^ote d'Azur, Observatoire de la C\^ote d'Azur, Laboratoire Lagrange, CNRS UMR 7293
CS 34229 06304 Nice Cedex 04 - France}

\author{Donatella Romano}
\affiliation{INAF - Osservatorio di Astrofisica e Scienza dello Spazio di Bologna, Via Gobetti 93/3 I-40129 Bologna, Italy}

\author{Luca Sbordone} 
\affiliation{European Southern Observatory, Alonso de Cordova 3107 Vitacura, Santiago de Chile, Chile}

\author{Rodolfo Smiljanic}
\affiliation{Nicolaus Copernicus Astronomical Center, Polish Academy of Sciences, ul. Bartycka 18, 00-716, Warsaw, Poland}

\author{Gra{\v z}ina Tautvai{\v s}ien{\. e}}
\affiliation{Institute of Theoretical Physics and Astronomy, Vilnius University, Sauletekio av. 3, 10258, Vilnius Lithuania}

\author{Mathieu Van der Swaelmen}
\affiliation{INAF - Osservatorio Astrofisico di Arcetri, Largo E. Fermi 5, Florence  I-50125,  Italy}

 \author{Elisa Delgado Mena}
 \affiliation{Centro de Astrofísica da Universidade do Porto, Rua das Estrelas, 4150-762 Porto, Portugal}

\author{Gerry Gilmore}
\affiliation{Institute of Astronomy, University of Cambridge, Madingley Road, Cambridge CB3 0HA, United Kingdom}

\author{Sofia Randich}
\affiliation{INAF - Osservatorio Astrofisico di Arcetri, Largo E. Fermi 5, 50125, Florence, Italy}

\author{Giovanni Carraro}   
\affiliation{Dipartimento di Fisica e Astronomia, Universit\`a di Padova, Vicolo dell'Osservatorio 3,  Padova,   I-35122, Italy}

\author{Anna Hourihane}
\affiliation{Institute of Astronomy, University of Cambridge, Madingley Road, Cambridge CB3 0HA, United Kingdom}

\author{Laura Magrini} 
\affiliation{INAF - Osservatorio Astrofisico di Arcetri, Largo E. Fermi 5, Florence  I-50125,  Italy}

\author{Lorenzo Morbidelli} 
\affiliation{INAF - Osservatorio Astrofisico di Arcetri, Largo E. Fermi 5, Florence  I-50125,  Italy}

\author{S\'ergio Sousa} 
\affiliation{Instituto de Astrof\'isica e Ci\^encias do Espa\c{c}o, Universidade do Porto, CAUP, Rua das Estrelas, 4150-762 Porto, Portugal}

\author{C. Clare Worley} 
\affiliation{Institute of Astronomy, University of Cambridge, Madingley Road, Cambridge CB3 0HA, United Kingdom}        




\begin{abstract}
This paper focuses on  carbon that is one of the most abundant elements in the Universe and is of high importance in the field of  nucleosynthesis and galactic and stellar evolution. Even  nowadays,  the  origin  of  carbon   and  the  relative  importance   of  massive  and  low-  to intermediate-mass stars in producing it is still a matter of debate. In this paper we aim at better understanding the origin of carbon by studying the trends of [C/H], [C/Fe],and [C/Mg] versus [Fe/H], and [Mg/H] for  2133 FGK dwarf stars from the fifth Gaia-ESO Survey internal data release (GES iDR5). The availability of accurate parallaxes and proper motions from Gaia DR2 and radial velocities from GES iDR5 allows us to compute Galactic velocities, orbits and absolute magnitudes and, for 1751 stars,  ages via a Bayesian approach. Three different selection methodologies have been adopted to discriminate between thin and thick disk stars.  In all the cases, the two stellar groups show different abundance ratios, [C/H], [C/Fe], and [C/Mg], and span different age intervals, with the thick disk stars being, on  average, older than those in the thin disk.
The behaviours of [C/H], [C/Fe], and [C/Mg] versus [Fe/H], [Mg/H], and age all suggest that C is primarily produced in massive stars like Mg. The increase of [C/Mg] for young thin disk stars indicates  
a contribution from low--mass stars or the increased C production from massive stars  at high metallicities 
due to the enhanced mass loss.
The analysis of the orbital parameters $R_{\rm med}$ and $|Z_{\rm max}|$ support an ``inside--out'' and ``upside--down'' formation scenario for the disks of Milky Way.

\end{abstract}



\section{Introduction} \label{sec:intro}

The present chemical composition of stars in the Milky Way (with the exception of hydrogen, helium, and  trace amounts of lithium and beryllium produced during the formation of the universe) comprises various nucleosynthetic products forged 
in previous generations of stars. Since the production history of each element can follow different nucleosynthesis pathways (probing different astrophysical 
processes, sites, timescales, and/or stellar-progenitor masses), all of the elements play potentially important roles in our understanding of Galactic chemical evolution (GCE).  
However, in this work, we focus on carbon  that next to hydrogen, helium (actually linked to the Big Bang) and oxygen, is the most abundant element 
in the Universe and is of  high importance in the field of galactic nucleosynthesis, stellar evolution,  exoplanets, and astrobiology.

Even nowadays, the origin of carbon is somewhat uncertain and  the relative importance of massive and low- to intermediate-mass stars is still a matter of debate.
\citet{GUS99} conclude that carbon is mainly contributed from super-winds of metal-rich massive stars, and not from low-mass stars. 
\citet{CHI03b}, \citet{MAT03}, \citet{CHI03a},  \citet{BEN06}, and \citet{MAT10} find strong indications that carbon is produced in low- and intermediate-mass stars. \citet{SHI02} find that
carbon is contributed by super winds of metal-rich massive (M$>$8\,M$_{\Sun}$) stars in the early stages of disk formation in the Galaxy,
 while a significant amount of carbon is contributed by low-mass stars in later stages. Other works in the literature favour different mixtures between the relative importance of  massive and low- to intermediate-mass stars
\citep[e.g.][]{LIA01,AKE04,GAV05}  
 while others \citep[e.g.][]{CAR05,HEN00} suggest massive stars as the main
carbon source. In any case, it is important to notice that the relative contributions from low- to intermediate--mass stars and massive stars depend strongly on the age and past evolutionary rate of the stellar system that is being scrutinised, hence, the conclusions drawn for the solar vicinity do not necessarily hold for any other system and/or Galactic region \citep{CAR05, ROM19}.
 

Different views on the relative role of high-mass and intermediate- to low-mass stars have reflected uncertainties
on the carbon production,  the dredge-up and rotation effects, stellar yields and metallicity-dependent mass loss from stars 
of different mass and chemical composition \citep[e.g.][]{MEY02,VAN97}. Additionally, significant uncertainties on the observed carbon abundances have hindered the checks of compatibility between
theoretical results and those obtained empirically. 
Nucleosynthesis and Galactic evolution of C can be studied by determining its abundance mainly in main-sequence stars (with different ages and metallicities)
of spectral types F, G, and K because their atmospheres still present essentially the original chemical composition of their birth
sites. However, there is still considerable uncertainty about the abundances of C  because of inaccuracy of oscillator strengths (log\,{\it gf}),
incompleteness of the available atomic and molecular lines, dependence of results from non-LTE corrections and atmospheric models 
applied \citep{ASP05,AMA19a,AMA19b}. The spectral lines invoked to spectroscopically determine the abundance
of carbon, such as atomic lines (CI) or molecular lines (CH, C$_2$, CO)  have different characteristics, depending on the type of stars, in terms of the sensitivity to  3D and/or non-LTE effect.

So far, carbon abundance determination from high resolution spectra of F and G main-sequence stars in the solar neighbourhood have not provided consistent results.
\citet{RED06} and \citet{NIS14} have found evidence of a systematic difference in the trend of [C/Fe] vs [Fe/H]  between thin- and thick-disk stars. 
The work of \citet{RED06} work is based on a sample of about 200 thin and 100 thick disk stars whose membership to the thin or thick disk is based on their kinematics
while \citet{NIS14} use two smaller samples of stars (57 thin  and 25 thick disk stars, respectively)  selected using the [$\alpha$/Fe]-[Fe/H] diagram as described by \citet{ADI13}.
In disagreement with the results of these studies, \citet{BEN06} show  almost flat and totally merged trends of [C/Fe] below [Fe/H] $\approx$-0.2\,dex for thin and thick disk stars.
Their work is based on the analysis of the [CI] line at 8727\,\AA, which is almost not affected by NLTE effects \citep[see][]{AMA19c}, for a small sample of  35 thin and 16 thick disk stars selected by using their kinematics and observed with high Signal to Noise Ratio (SNR$>300$) and high resolution (R$\sim 220,000$).

Since the trend of  [C/H]  with time or metal abundance [Fe/H] is presently not well constrained by stellar and galactic evolution models, 
much more insight should be gained from observations. 
With the advent of large spectroscopic surveys, such as Gaia-ESO  \citep[GES,][ESO programmes 188.B-3002
and 193.B-0936]{GIL12, RAN13}, APOGEE \citep{MAJ17}, and GALAH \citep{DES15}, it is  now possible to investigate, by using sizeable statistical samples, the behaviour of carbon in  different populations of our
Galaxy. Among several ongoing spectroscopic surveys, GES has provided high resolution spectra of stars belonging to various stellar populations of our Galaxy using the spectrograph FLAMES@VLT
\citep{PAS02}. GES aims at homogeneously deriving
stellar parameters and abundances in a large variety of environments, including the major Galactic components (thin and thick
disks, halo, and bulge), open and globular clusters, and calibration samples. The higher resolution spectra obtained with UVES \citep{DEK00}
allow the determination of the abundances for more than 30 different elements, including carbon.

In this paper we  derive  additional information on carbon abundances based on the trends of [C/H], [C/Fe], and [C/Mg] versus [Fe/H], [Mg/H], and age for stars in the Galactic disks by using a large sample of 2133 FGK dwarf stars whose spectra were extracted from 
the fifth Gaia-ESO Survey  internal data release (GES iDR5).  
We decided to use only dwarf stars and exclude red giants in which the atmospheric carbon and nitrogen abundances  are affected by internal mixing of material during the first
dredge-up phase \citep{IBE65}. This material from the core has
been enriched through the CN cycle, which results in a build up of
N and a depletion of C. The depth of the convective zone during the
first dredge-up phase is dependent on the mass of the star, resulting
in decreasing atmospheric C to N ratios in red giants
\citep[see][for a detailed explanation]{SAL15}.

Most of the stars in our sample have accurate parallaxes and proper motions available from the second Gaia data release  \citep[DR2][]{GaiDR2}.
Gaia DR2 database provides  a high-precision parallax and proper motion catalogue for over 1 billion sources, supplemented by precise
and homogeneous multi-band all-sky photometry and a large
radial velocity survey at the bright (G $\lesssim$ 13) end. The availability of precise fundamental astrophysical information required
to map and understand the Milky Way is thus expanded to a very substantial fraction of the volume of our Galaxy, well
beyond the immediate solar neighbourhood.  
The knowledge of accurate parallaxes and proper motions from Gaia DR2 and radial velocities from GES iDR5 allows us to compute Galactic velocities, orbits and absolute magnitudes  for 1804 stars   and then, by using Bayesian approach, ages for 1751 of them. In Section\,\ref{sec:sample} we present the starting stellar sample as obtained from GES iDR5 and the determination of carbon abundance. In Section\,\ref{sec:populations} we define the samples of thin and thick disk stars using three different selection methodologies. In Section\,\ref{sec:kinematics} and \ref{sec:C_result} we discuss the kinematical and chemical properties of the selected samples. Section\,\ref{sec:age} presents the trends of [C/H], [C/Fe], [C/Mg], $R_{\rm med}$, and $|Z_{\rm max}|$ with stellar ages for the thin and the thick disk stars, while conclusions are given in Section\,\ref{sec:conclusion}.

\section{The UVES-U580 stellar sample definition}\label{sec:sample}
We used the GES iDR5 internal release to extract the observed spectra of all the FGK dwarf stars 
obtained with the UVES spectrograph in a setup centred at 580\,nm (hereafter UVES-U580 sample) at a resolution $R\sim$47,000.
The spectra were exposed onto two CCDs, resulting in a wavelength coverage of 4700-6840\,\AA~with a gap of about 50\,\AA~in the centre.
Data reduction of the UVES spectra has been performed using a workflow specifically developed for this project \citep{SAC14}.
The GES iDR5 release contains, together with the stacked spectra and the tables of metadata summarising these spectra, 
also  radial ($V_{\rm r}$), and rotational velocities ($v\sin i$), recommended stellar atmosphere parameters (effective temperature, $T_{\rm eff}$, surface gravity, log\,$g$,
iron abundance, [Fe/H], and  microtubulence, $\xi$), and individual element abundances\footnote{All abundances of element X are given in the following format:
$\log \epsilon_{X}  = \log \frac{N_X}{N_H} + 12.0$} including carbon. 
The UVES-580 spectra were analyzed with the Gaia-ESO multiple pipelines strategy, as described in \citet{SMI14}. The results of each pipeline are combined 
with an updated methodology  to define the final set of recommended values of the atmospheric parameters and chemical 
abundances that are part of GES iDR5 \citep[see also][]{MAG17, HOU19}. Average uncertainties in the atmospheric parameters are 55 K, 0.13 dex,
and 0.07 dex for $T_{\rm eff}$, log\,$g$, and [Fe/H], respectively \citep{MAG18}. 
\begin{figure}[htbp]
\includegraphics[width=0.8\textwidth]{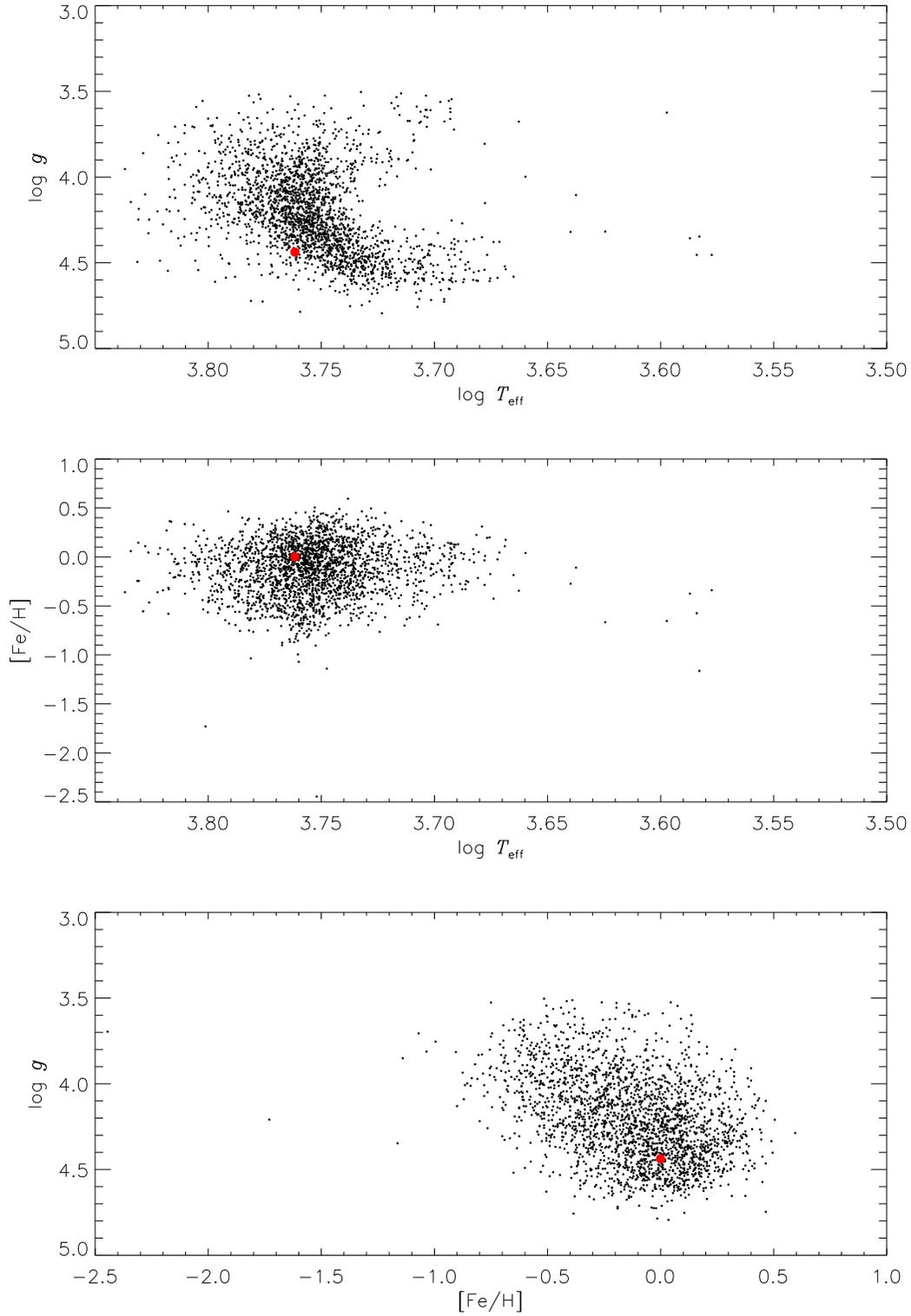}
\caption{GES iDR5 atmospheric parameters of the original sample of 2261 dwarf stars (black points) with superimposed the Sun position (red cicle):  log\,$g$ vs log\,$T_{\rm eff}$ (upper panel);
 [Fe/H] vs  log\,$T_{\rm eff}$ (middle panel); log\,$g$ vs [Fe/H] (lower panel).}
\label{fig:param}
\end{figure}

\subsection{Stellar parameters}
A first sample of dwarf stars was obtained by performing a Structured Query Language (SQL) search to select all the
stars in the 3750-7000\,K and 3.50-5.00\,dex  effective temperature  and surface gravity ranges\footnote{Effective temperature and surface 
gravity ranges approximately covered by F,G, and K dwarf stars.} 
observed with U580 setup and characterized  by an  SNR  greater than 10. 
Then,  we removed all the stars with some peculiarity or binarity flag,  lack of error estimates of the stellar atmosphere parameter values, or without iron abundances from Fe\,I or Fe\,II lines getting a sample of 2261 stars.
In such a way we obtained a sample well suitable for our analysis since it contains objects with homogeneously determined  
$T_{\rm eff}$, log\,$g$, [Fe/H], and detailed chemical composition, spanning the following ranges: 
$T_{\rm eff}$  from 3779 to 6868\,K; log\,$g$ from 3.50 to 4.80\,dex; [Fe/H] from -2.44 to +0.60\,dex. 
The atmospheric parameter coverage is shown in Figure\,\ref{fig:param}. In the following, to work with a statistical significant data--set, we limit our analysis to stars with $T_{\rm eff}>4450$\,K, and [Fe/H]$>$-1.0\,dex, thus reducing the sample to a total of 2248 stars.

\subsection{Carbon abundances}
Out of the 2248 stars of our sample 1936 have an estimate of carbon abundance from atomic lines in GES iDR5.   The abundance determination of C is quite 
challenging and the values of C/H derived by GES iDR5 are, in general, less accurate than the corresponding values for the 
other elements. In particular, the estimated ``C1'' GES iDR5 carbon abundance for the stars in our sample is based on the analysis of only 2 
(1500 stars) or even 1 (436 stars) spectral lines.  
The ``C1Err'' uncertainties  are estimated considering the errors 
on the atmospheric parameters, random errors (mainly caused by uncertainties of the continuum placement and by the signal-to-noise) and span a range from 0.01 to 0.58\,dex with the bulk of data at 0.05\,dex.

It is worthwhile to point out that most of the C/H values were derived by using synthetic spectra computed from MARCS atmosphere models without full consistency between the chemical
composition used to build the atmosphere structure and that one actually used in synthesizing the emergent spectrum. In fact, even if MARCS models use the Opacity Sampling (OS) method, GES WGs adopted grids of models with fixed chemical composition to derive the stellar abundances without  any iterative procedure, i.e without injecting the derived abundances in the atmosphere models and recomputing the atmosphere structure and abundances until consistency is achieved. 
Such an inconsistency
may introduce systematic errors in the abundance determination, in particular when dealing with elements like carbon which may affect
significantly the overall opacity. Therefore, in order to remove such an uncertainty and with the goal of increasing the number of stars in our sample with 
determined C/H, we decided to re-analyze all their spectra.

\begin{figure}[htbp]
\includegraphics[width=0.95\textwidth]{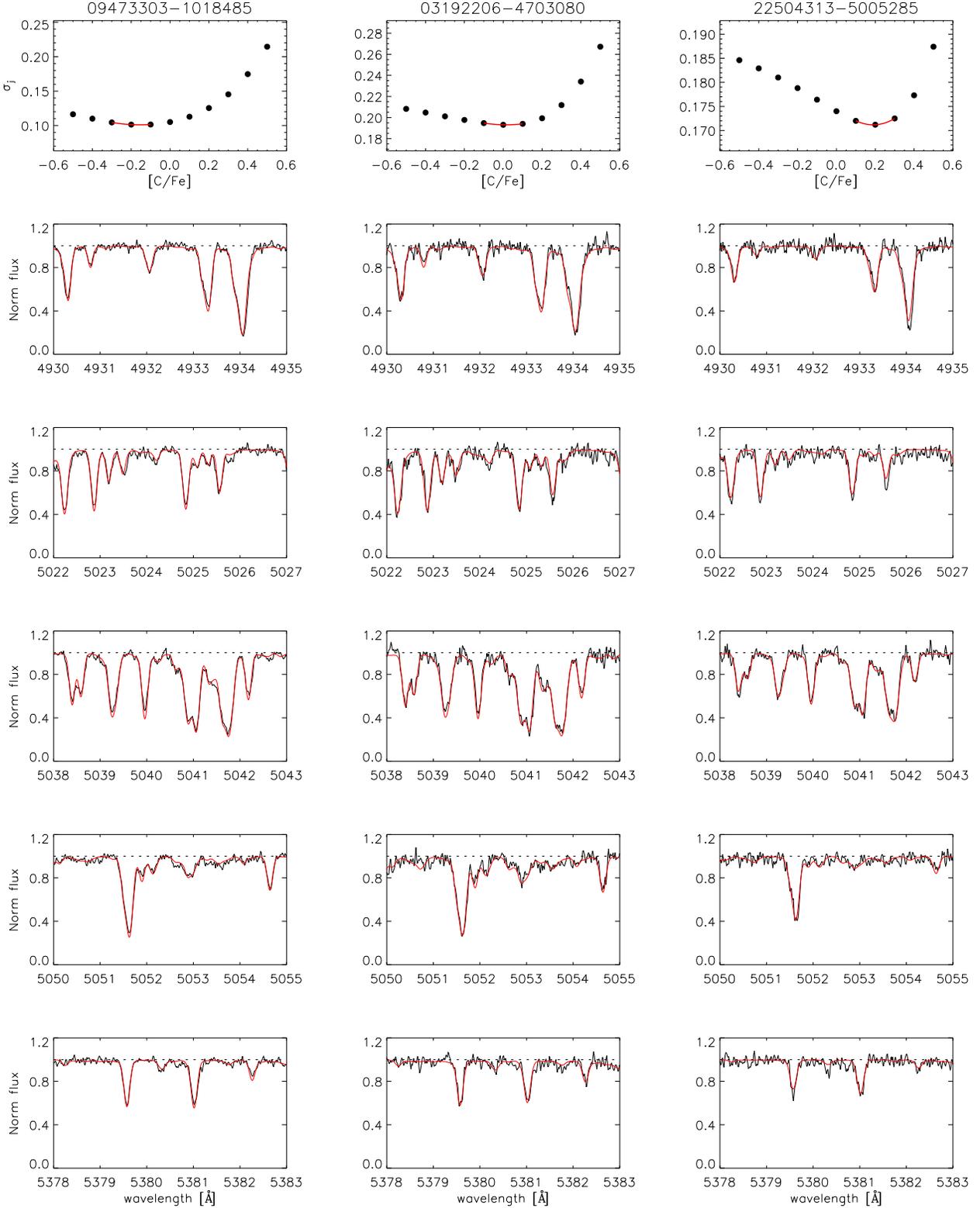}
\caption{Examples of determination of [C/Fe] for 3 stars. Top panels: trends of $\sigma_j$ vs [C/Fe] (black dots) and parabolic interpolations (red lines);
other panels: comparison between observed spectra (black lines) and corresponding synthetic ones (red lines) computed with the ``best'' [C/Fe] (see text).
\label{fig:min_sig}}
\end{figure}

\begin{figure}[ht!]
\includegraphics[width=0.9\textwidth]{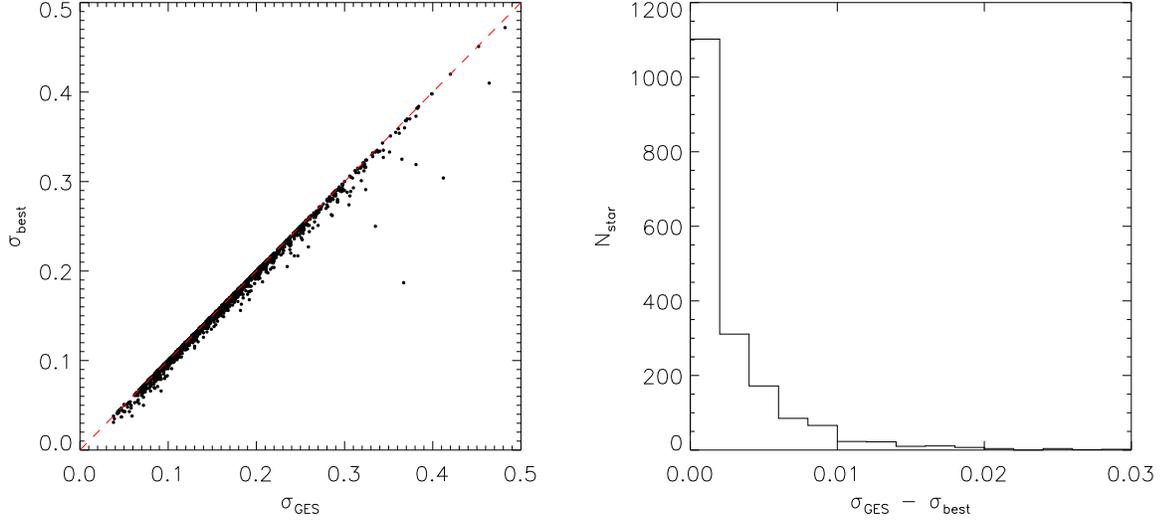}
\caption{Comparison between  $\sigma_{\rm best}^i$ and $\sigma_{\rm GES}^i$ for the  1870 stars with [C/Fe] from this paper and from GES iDR5 (left panel) and distribution of the differences $\sigma_{\rm GES}^i - \sigma_{\rm best}^i$. Eleven stars with differences greater than 0.03  were excluded from the distribution plot to increase readability. \label{fig:sigma}}
\end{figure}

\begin{figure}[htbp]
\includegraphics[width=0.55\textwidth]{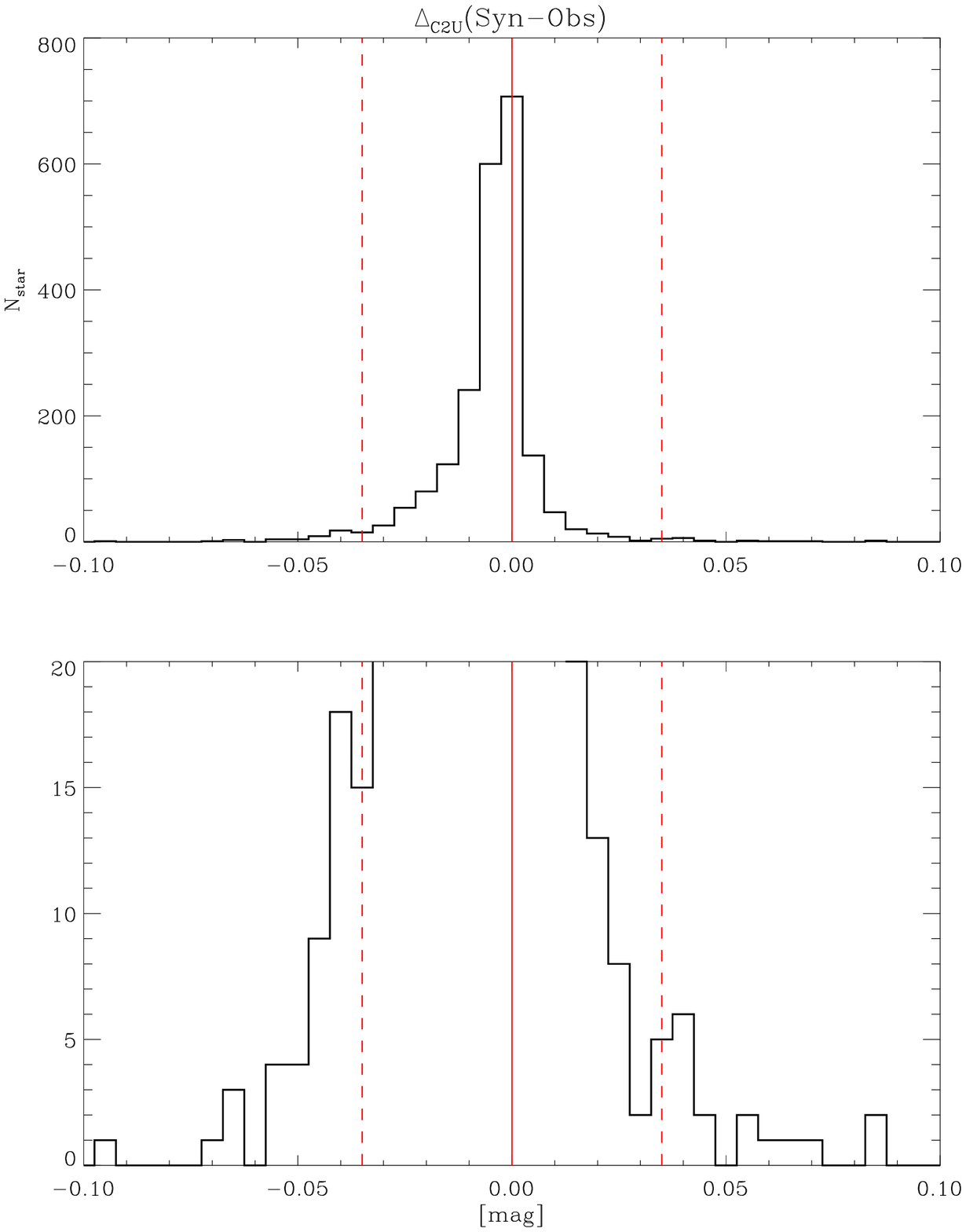}
\caption{Distribution of the differences between C2U indices derived for each
pair of ``best'' [C/Fe] synthetic and UVES-U580 spectra; the lower panel zooms the y-scale
and shows the presence of few stars with $\Delta _{\rm C2U}$ values outside 
 $3\sigma_{\Delta {\rm C2U}}$ (red dashed lines).
\label{fig:C2U}}
\end{figure}

\subsubsection{Model atmosphere and synthetic spectra} \label{sec:synthetic}
To estimate  C/H abundances we used the stellar atmosphere  ATLAS12  code  \citep{KU05a} and the spectral synthesis program SPECTRUM v2.76f \citep{GRA94}  to compute 
for each of the 2248 stars, assuming different C/H values,  its model atmosphere and theoretical spectrum, respectively. 

We used ATLAS12 since it allows us to generate ad--hoc atmospheric models for any individual element chemical composition and microturbulence parameter ($\xi$), through the OS technique. As starting point, we adopted for the reference solar abundances those obtained by \citet{GRE07} which have a wide consensus in the literature and whose validity is also confirmed, 
within the quoted uncertainties, by the abundance determinations derived by the Working Group 11 (WG11) of the Gaia-ESO consortium \citep{MAG17} 
from the analysis of   UVES spectra of the Sun and M67 giant stars obtained with the U580 and U520 setups\footnote{https://www.eso.org/sci/facilities/paranal/instruments/uves/doc.html}. 
Then, for each {\it i}-th star, we used its GES iDR5 atmospheric parameter values ($T_{\rm eff}$, log\,$g$,[Fe/H], and  $\xi$) and individual element abundances but for C (for those elements with no estimate of [X/Fe] we assumed [X/Fe]=0).
For each star the  atmosphere model was calculated   starting from the closest model, in  the atmosphere parameter space,  among those used for calculating
the INTRIGOSS  high resolution synthetic spectral library \citep{FRA18}.
Its convergence was checked according to the convergence criteria recommended in the ATLAS cookbook\footnote{http://atmos.obspm.fr/index.php/documentation}. 
In general, a model is accepted if, at the end of the computing iteration, the flux and the flux derivative errors are, for each layer, below 1\% and 10\%, respectively.
Only for a few models (among the coldest)
we needed to significantly increase the number of iterations from the standard figure of 25 to reach the convergence.
Eventually, a further check on the reliability of the new obtained ATLAS12 models was done by looking, for each of them, at the behaviours  
of temperature, gas pressure, electron density, Rosseland absorption coefficient, and radiation 
pressure  at all Rosseland optical  depths.

Then, to obtain  the corresponding
emergent flux and normalized spectrum, we used
SPECTRUM v2.76f. The SPECTRUM  code calculates an LTE synthetic spectrum starting from  a given model atmosphere. The code
also requires a line list of atomic and molecular
transitions and we used the INTRIGOSS line list whose accuracy was established in \citet{FRA18}\footnote{http://archives.ia2.inaf.it/intrigoss/}.
 SPECTRUM was used to deliver both the
stellar-disk-integrated normalized spectrum, S$_{\rm N}$, and the absolute
monochromatic flux at the stellar surface, S$_{\rm F}$,  in the spectral range 4830-5400\,\AA. In particular, we computed for each star a set of  11  models and synthetic spectra with  [C/Fe]$_j=-0.5+(j-1)\times0.1$\,dex (with {\it j}=1,..,11) and  the  model and synthetic spectrum, S$^{\rm i,GES}_{\rm N}$, computed at the GES iDR5 [C/Fe] ratio when available.
The use of ATLAS12  and SPECTRUM v2.76f  codes which allowed us to specify the same microturbulence and  individual  element abundances both in 
deriving the atmosphere structure and the synthetic spectrum  guarantees the full consistency between atmosphere models and synthetic spectra. 
Eventually, since 
the synthetic spectra were computed at a resolving power $R\sim240,000$, they were broadened by using the GES iDR5  $v\sin i$ stellar values and degraded at the resolution of UVES spectra ($R\sim$47,000).

In order to remove the instrumental signature in the observed (stacked) UVES-U580 spectra we used, for each star {\it i}, the {\it j} normalized synthetic S$^{\rm i,j}_{\rm N}$ spectra  
to obtain from the corresponding observed  UVES-U580 one a set of normalized observed  spectra (O$^{\rm i,j}_{\rm N}$). 
The normalization was performed by applying the technique described in \citet{FRA18}. We searched for quasi--continuum flux reference points in S$^{\rm i,j}_{\rm N}$
(i.e. wavelength points with flux levels in excess of 0.95) and we used the same points in the corresponding observed UVES spectrum to derive the continuum shape via a low-order polynomial fitting
of the ratio between  observed  and  synthetic spectra. 
Eventually, the observed spectrum is divided by the so computed polynomial to obtain the normalized spectrum O$^{\rm i,j}_{\rm N}$. 
With this technique of matching 
the continuum levels of observed and corresponding synthetic spectra, we also obtained the observed ``flux calibrated'' spectra  O$^{\rm i,j}_{\rm F}$ 
by using the ratio between  observed UVES spectrum and the corresponding S$^{\rm i,j}_{\rm F}$ in the same 
reference points previously defined via S$^{\rm i,j}_{\rm N}$.

\subsubsection{{\rm [C/Fe]} determination}
\label{sec:CFe_det}
 For  carbon abundance determination  we used a spectrum synthesis technique.
 We identify 5 wavelength regions listed in Table\,\ref{tab:C_lines},  characterized by higher sensitivity to C abundance because of the presence of relatively strong C\,I lines. We look at each {\it i} star and, for each {\it j} pair of spectra, i.e. for different [C/Fe] values, we computed the total standard deviation ($\sigma_j^i$) between O$^{\rm i,j}_{\rm N}$ and S$^{\rm i,j}_{\rm N}$ in these wavelength regions. Then, using a parabolic fitting, we determine the ``best'' [C/Fe] value corresponding to the position of the minimum (if any) of $\sigma_j^i$ vs [C/Fe].
 
First of all, we fine--tuned the log\,{\it gf} of the CI lines  by using the same technique described in \citet{FRA18}, i.e. by comparing the synthetic solar spectrum with an observed one with SNR$\sim 4000$ after degrading both of them at the UVES--U580 resolution. The so obtained, and used in our analysis, log\,{\it gf} (15 lines did not need any log\,{\it gf} tuning) are reported in Table\,\ref{tab:C_lines} and we point out that our  astrophysical log\,{\it gf} values for the only two of our lines which are flagged ``recommended (Y)'' in the GES database are in good agreement with those reported in  the Table ``LineList'' of GES iDR5.
Then the procedure above--described  was applied to the solar spectra and led to  [C/Fe]=0.01\,dex for the Sun thus assessing the absence of a systematic offset in the derived [C/Fe] values.   

Eventually,  we were able  to obtain for 2133  stars fiducial [C/Fe] values  whose uncertainties  we fixed, to be conservative, at $\pm 0.05$\,dex, i.e. half-step in our [C/Fe] grid of models and synthetic spectra.
No clear minimum in $\sigma_j^i$  was detected for the other 115 stars, thus preventing a sound determination of [C/Fe]. 
Some examples of the adopted procedure are shown in Figure\,\ref{fig:min_sig}. 

In order to double check the derived carbon abundances, we computed, on the basis of these 
updated [C/Fe] values, new (hereafter ``best'') model atmospheres and synthetic spectra, S$^{\rm i,best}_{\rm N}$ 
and  S$^{\rm i,best}_{\rm F}$, for each of the 2133 stars and compared them  with the observed spectra in the five wavelength regions sensitive to C abundance. The validity of the  [C/Fe] values was confirmed by the good agreement between observed and synthetic spectra (see  Figure\,\ref{fig:min_sig}).
We also computed the $\sigma_{\rm best}^i$'s and the $\sigma_{\rm GES}^i$'s for  the 1870 stars which have both our and GES iDR5 estimate of C/H.  Figure\,\ref{fig:sigma} shows that S$^{\rm i,best}_{\rm N}$ spectra reproduce as well as the S$^{\rm i,GES}_{\rm N}$ the observed spectra for  1143 stars ($\sigma_{\rm GES}^i-\sigma_{\rm best}^i < 0.002$). For the other 727 stars $\sigma_{\rm best}^i$ is smaller than $\sigma_{\rm GES}^i$ indicating that our  estimates of C/H are more accurate than those reported in the  GES iDR5 ``RecommendedAstroAnalysis'' Table. 

\begin{figure}[htbp]
\includegraphics[width=0.50\textwidth]{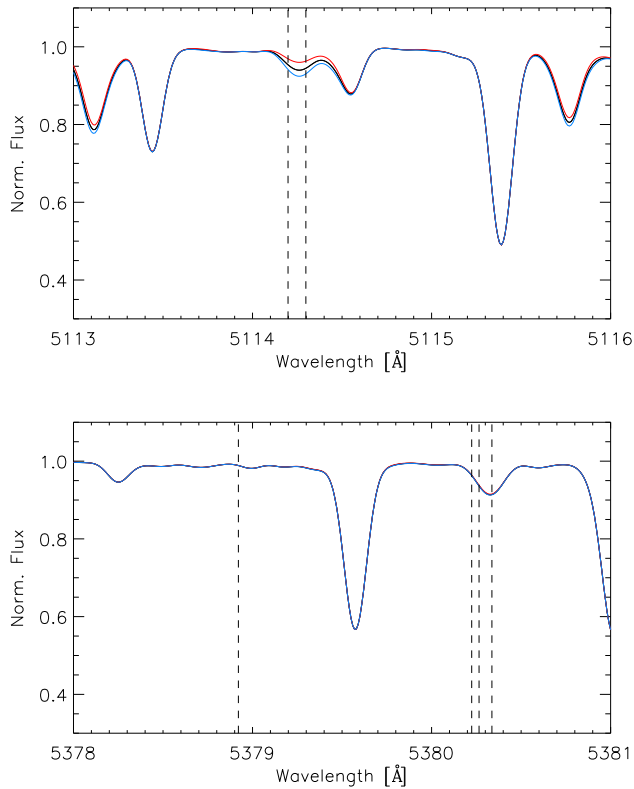}
\caption{Comparison between the three synthetic spectra computed for the star 09473303-1018485  by using the same [C/Fe]=-0.21 and three different [O/Fe] values, namely [O/Fe]=-0.4 (blue line), 0.0 (black line), and +0.4 (red line). Upper panel shows a region with two strong C$_2$ lines (whose positions are indicated by the vertical dashed lines); lower panel shows one of the five regions containing the CI lines used to derive [C/Fe] (the vertical dashed lines indicate the positions of the four CI lines listed in Table\,\ref{tab:C_lines}).
\label{fig:C2_C}}
\end{figure}

Since the UVES-U580 spectra contain the C$_2$ bands of the Swan system \citep{SWA57} and, in particular,  the one used by \citet[Table 2]{GON16} 
to define the C2U index (bandpass feature: 5087--5167\,\AA, bandpass ``continuum'': 5187--5267\,\AA), we decided to use the C2U index to further check our [C/Fe] determinations.
In such a way we will use C$_2$ molecular features to verify our C abundances which are based on atomic lines. We computed for the 2133 stars the C2U index from both S$^{\rm i,best}_{\rm F}$ and   O$^{\rm i,best}_{\rm F}$ obtained as described in Section\,\ref{sec:synthetic}.
Figure\,\ref{fig:C2U} shows the distribution of the differences between the synthetic and the observational C2U index values. As can be seen,
in most cases the differences are within $3\sigma_{\Delta {\rm C2U}}=0.036$\,mag (there are only 66 stars, 3\,\% of the total, that show larger differences). By comparing this figure with Figure\,5 in \citet{FRA18} it can be noticed that the outliers with $\Delta_{\rm C2U}$ larger, in absolute value, than 0.1\,mag have disappeared and that the bulk of data is now within $\pm 0.036$ instead of  $\pm 0.05$. The largest differences are for the stars with [Fe/H]$\gtrsim$+0.2\,dex. We recall that the strength of C$_2$ lines depends not only on carbon abundance but also, indirectly, on nitrogen and oxygen abundances because of the competing role of CN and CO.  
In Figure\,\ref{fig:C2_C} we show the effect, on C$_2$ and CI lines, of different oxygen abundances in the S$_{\rm N}^{\rm best}$ of the star 09473303-1018485 (also shown in Figure\,\ref{fig:min_sig}). Three synthetic spectra were computed by using [O/Fe]=-0.4, 0.0, and +0.4 with the same [C/Fe]=-0.21. As can be seen, the strength of the C$_2$ lines (upper panel) is different for the different [O/Fe] values, i.e. as expected lower [O/Fe] lead to  stronger C$_2$ lines. On the other hand the atomic CI lines (lower panel) are unaffected by the oxygen abundance (the three different synthetic spectra plotted with different colors practically coincide). Actually, the sensitivity of the C$_2$ lines to the oxygen abundance is the reason why we preferred to derive carbon abundances from atomic lines instead of using molecular ones even if atomic lines may be affected by NLTE (see discussion below).

The small differences shown in Figure\,\ref{fig:C2U}, indicate that, in any case, our estimates of [C/Fe] reproduce in an acceptable way not only the CI atomic lines but also the strength of the C$_2$ Swan bands. This fact, suggests that in our ``best'' models and synthetic spectra  we used acceptable estimates of 
N/H and O/H even if no detailed information on nitrogen  and oxygen abundance are  available in GES for our stars.

Recent studies in late-type stars have demonstrated the potentially large impact of 3D non-LTE effects on carbon abundance determination \citep[e.g.][]{AMA19c}.
The paper by \citet{AMA19b} shows that in the metallicity regime of our stars ([Fe/H]$>-1$) the 3D-NLTE corrections do not change significantly the [C/Fe] ratios  (see their Figure\,11). 
Their paper provides, in any case, a tool to correct 1D-LTE carbon abundances to take into account the 3D-NLTE effects. We used a code kindly provided us by A. M. Amarsi (private communication) to compute for all our stars the required 3D-NLTE corrections. It is worth noticing that our C/H estimates were not derived from equivalent width measurements and that \citep{AMA19b} provides corrections only for 2 CI lines given in Table \,\ref{tab:C_lines} thus compelling us to consider the corrections only as a first approximation. 
The corrections we computed are all between $\pm 0.05$\,dex i.e. of the order of our uncertainties. Moreover, we did not find any systematic difference in the corrections between the thin and thick disk samples defined in Section\,\ref{sec:populations} (for the actual values see Sections \ref{sec:Adibek}, \ref{sec:Bensby}, and \ref{sec:Orbits}). In the following we will use our atomic 1D-LTE C/H values for comparing overall trends since 3D-NLTE corrections do not affect significantly, at least in a first approximation,  our results.

\begin{deluxetable}{cccc}[htbp]
\tablecaption{Wavelength regions used to estimate [C/Fe] via comparison of synthetic and observed normalized spectra. \label{tab:C_lines}}
\tablecolumns{4}
\tablewidth{0pt}
\tablehead{
\colhead{wavelength region} &
\colhead{C\,I lines} &
\colhead{log {\it gf}} &
\colhead{References} \\
\colhead{\AA} &
\colhead{\AA} &
\colhead{} &
\colhead{} 
}
\startdata
4930 - 4935 & 4930.276 & -3.480 & \citet{KP} \\
          & 4932.039 & -1.684 & this work \\
          & 4934.301 & -4.930 & \citet{KP} \\
5022 - 5027 & 5023.849 & -2.400 & \citet{KUR95} \\
          & 5024.916 & -2.700  & \citet{KUR95} \\
5038 - 5043 & 5039.057 & -2.200 & this work \\
          & 5039.100 & -2.286 & \citet{MWRB} \\
          & 5039.919 & -3.940 & \citet{KUR95} \\
          & 5040.134 & -2.500 & \citet{KUR95} \\
          & 5040.765 & -2.600 & \citet{KUR95} \\
          & 5041.481 & -1.700 & \citet{KUR95} \\
          & 5041.796 & -2.500 & \citet{KUR95} \\
5050 - 5055 & 5051.579 & -2.480 & \citet{KP} \\
          & 5052.142  & -1.303 & this work \\
          & 5053.515 & -1.555 & this work \\
          & 5054.619 & -3.690 & \citet{KP} \\
5378 - 5383 & 5378.921 & -4.640 & \citet{KP} \\
          & 5380.224  & -2.030 & \citet{KP} \\
          & 5380.265  & -2.820 & \citet{KP} \\
          & 5380.312 & -1.692 & this work \\
\enddata
\end{deluxetable}

\section{Thin and thick disk samples}\label{sec:populations}

In this paper we want to investigate if there is any difference in the C abundance behaviour in stars belonging to the thin or to the thick Galactic disk. 
To achieve this goal we need
to select the stars in our sample which are part of each disk component.   Out of the several approaches adopted in the literature
for identifying thin and thick disk stars the two most used  are those based on  purely
a chemical  or a kinematical   approach even if combinations of kinematics, metallicities, and stellar ages can also  be adopted \citep[see for example][]{FUH98}. 

\begin{figure}[htbp]
\includegraphics[width=0.70\textwidth]{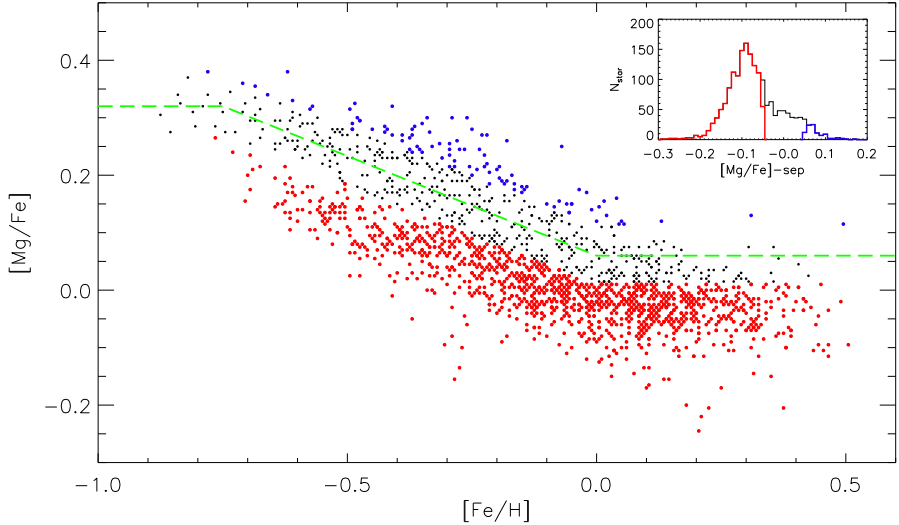}
\caption{[Mg/Fe]--[Fe/H] diagram used to chemically select thin (red points) and thick (green points) disk stars (see text); stars falling in the avoidance region (see Section\,\ref{sec:Adibek}) are indicated by black points.  On the top-right corner we show the distributions of [Mg/Fe] after subtracting the separation line (dashed green line in the main plot). 
\label{fig:MgFeAdib}}
\end{figure}

\subsection{Chemical selection}
\label{sec:Adibek}
Among  several methods proposed so far to differentiate the Galactic disks, the ones based on stellar abundances 
have extensively been considered robust, mainly  because chemistry is a relatively stable property 
of stars \citep[][and references therein]{ADI13}. 
Following such approach we use the position of the stars in the [Mg/Fe]--[Fe/H] plane to discriminate those belonging to the two different  disks. 
Our analysis differ from that one by \citet{ADI13} since we used Mg, as in \citet[][]{KOR15}, instead of an average of the abundances of  $\alpha$ elements (i.e. Mg, Si, and Ti). Our choice is based on the fact that the different $\alpha$-elements are produced by different stellar progenitors.
In particular, Si and Ti are produced also by Type\,Ia\,SNe whereas Mg is not \citep{CES07, ROM10}.  Figure\,\ref{fig:MgFeAdib} shows the separation plot we adopted. The separation line (dashed green line) is somewhat arbitrary and was obtained in analogy with those used  in \citet{ADI11,ADI13,HAY13}. We classify as thick disk stars (blue points) those above the separation line and as thin disk stars (red points) those below. To minimize contamination we adopted an avoidance region  of $\pm 0.05$\,dex.
In such a way, we obtained two samples of 1267 thin disk stars (Thin$^{\rm C}$ sample, where the superscript ``C'' recall that was obtained via a Chemical selection) and 99 thick disk stars (Thick$^{\rm C}$ sample). On the top-right corner of Figure\,\ref{fig:MgFeAdib} we show 
the distributions of [Mg/Fe] after subtracting the separation line for all the stars (in black) and for the two samples (in red and in blue).  The average  3D-NLTE corrections for the Thin$^{\rm C}$  and Thick$^{\rm C}$ samples are $-0.02\pm 0.02$\,dex and $-0.03\pm 0.01$\,dex, respectively.

\subsection{Kinematical selection}
\label{sec:Bensby}
Among the different criteria to separate thin- and thick-disk populations  using kinematical properties of the stars, a popular strategy is based on the assumption of Gaussian velocity distributions in each Galactic component \citep{BEN03,BEN14}:

\begin{equation} \label{eq:fProb}
f(U,V,W)=k\cdot\exp\left(-\frac{(U_{\rm LSR}-U_{\rm asym})^2}{2\sigma_U^2}-\frac{(V_{\rm LSR}-V_{\rm asym})^2}{2\sigma_V^2}-\frac{W_{\rm LSR}^2}{2\sigma_W^2}\right)
\end{equation}

where

\begin{eqnarray}
     & k & =  \frac{1}{(2\pi)^{2/3}\sigma_U\sigma_V\sigma_W} \nonumber 
\end{eqnarray} 

normalises the expression,  $U_{\rm LSR}$ , $V_{\rm LSR}$ , and $W_{\rm LSR}$ are the Galactic velocities of the stars in the Local Standard of Rest (LSR), $\sigma_U$, $\sigma_V$, $\sigma_W$ are the characteristic velocity dispersions for the different populations, and $U_{\rm asym}$ and $V_{\rm asym}$
are the asymmetric drifts.  For a given star, when computing the likelihoods of belonging to one of the  Galactic populations (i.e. $P_{\rm Thin}$, $P_{\rm Thick}$, 
$P_{\rm Halo}$, $P_{\rm Hercules}$), one has to take into account the local number densities of each population 
($X_{\rm Thin}$, $X_{\rm Thick}$, $X_{\rm Halo}$, $X_{\rm Hercules}$):

 \begin{eqnarray} \label{eq:Prob}
     & P_{\rm Thin}~~~ & =  X_{\rm Thin}\cdot f_{\rm Thin} \nonumber \\
     & P_{\rm Thick}~~  & =   X_{\rm Thick}\cdot f_{\rm Thick} \\
        & P_{\rm Halo}~~~~  & =   X_{\rm Halo}\cdot f_{\rm Halo} \nonumber \\
        & P_{\rm Hercules}  & = X_{\rm Hercules}\cdot f_{\rm Hercules}  \nonumber 
  \end{eqnarray} 

\noindent where the sub--script Hercules refers to the ``Hercules'' stream \citep{FUX01}.

The values for the velocity dispersions, asymmetric drifts, and the observed fractions ($X$) of each population in the solar neighbourhood are given in \citet[][Table\,A.1]{BEN14}.

A shortcoming of this  approach is that the assumption of Gaussian distributions is valid only as a first-order approximation 
\citep{BIN10}. However, since there is no clear consensus  in the literature on the actual shape of the velocity distributions,  we used, for our purposes, normal distributions as also done, in several recent papers \citep[e.g.][]{BUD19}. 

\subsubsection{Space velocities and selection criteria}
\label{sec:Gal_Vel}
\begin{figure}[htbp]
\includegraphics[width=0.95\textwidth]{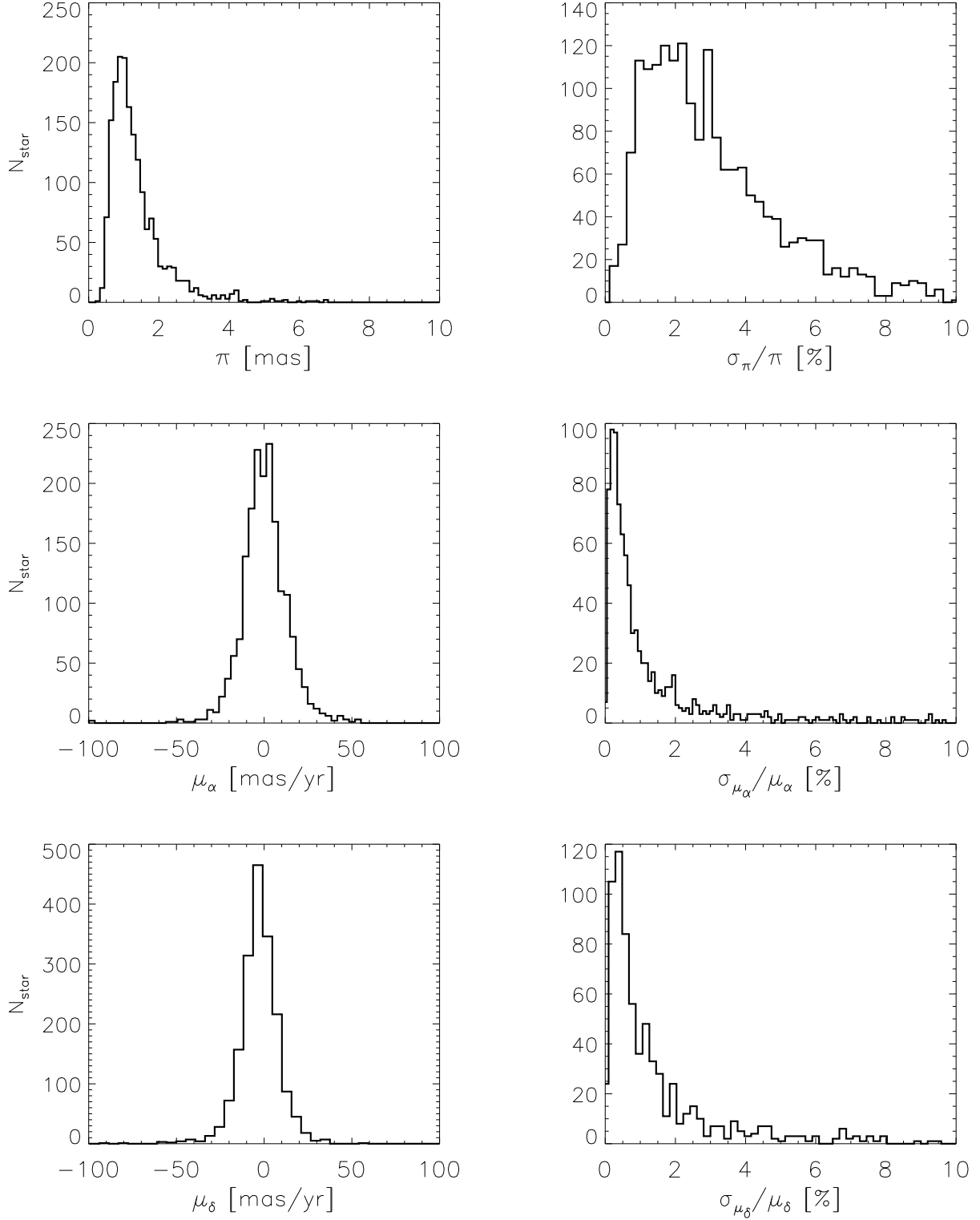}
\caption{Distributions  of $\pi, \mu_{\alpha}$, and $\mu_{\delta}$ (left panels) and their relative 
uncertainties (right panels) of the 1804 stars for which Galactic velocities are computed. Stars  with large parallaxes (18 stars with $\pi>10$\,mas), or large proper motions (10 stars with  $|\mu_{\alpha}|>100$\,mas/yr and 8 stars with  $|\mu_{\delta}|>100$\,mas/yr) were excluded from the figures to increase readability. \label{fig:ProperMotion}}
\end{figure}

In order to calculate space velocities ($U_{\rm LSR}$ , $V_{\rm LSR}$, and $W_{\rm LSR}$) for our sample stars, we need distances  (or parallaxes, $\pi$),  proper motions
($\mu_{\alpha}$, $\mu_{\delta}$) and radial velocities ($V_{\rm r}$). We searched for $\pi, \mu_{\alpha}$, and $\mu_{\delta}$  in
the second Gaia data release  (Gaia DR2,  \citealt{GaiDR2}) while for $V_{\rm r}$ we used  GES iDR5 values with typical percentage error below 7\%. For each star of our UVES-U580 sample we made a cross-match 
between the GES and Gaia DR2 coordinates by using  {\it gaiadr2.gaia\_source}  table and a match radius of 1\,arcsec. 
Each GES iDR5 star, if detected,  was associated to the nearest Gaia DR2 source leading to a sample of 2113 stars. Out of these stars we accepted only those characterized by small relative errors (less than 10\%) in $\pi$, $\mu_{\alpha}$, and $\mu_{\delta}$, thus obtaining an astrometric sample of 1804 dwarf stars  suitable for computing accurate Galactic velocities.   Figure \ref{fig:ProperMotion} show the distributions of  parallaxes (top-left panel), RA (middle-left panel) and DEC (bottom-left panel) proper motions,  and their percentage  errors (right panels) for the 1804 stars. As can be seen, the distributions of the relative errors in $\pi$, $\mu_{\alpha}$, and $\mu_{\delta}$ are peaked at about 2\%, 0.2\%, and 0.4\%, respectively.

\begin{figure}[htbp]
\includegraphics[width=0.5\textwidth]{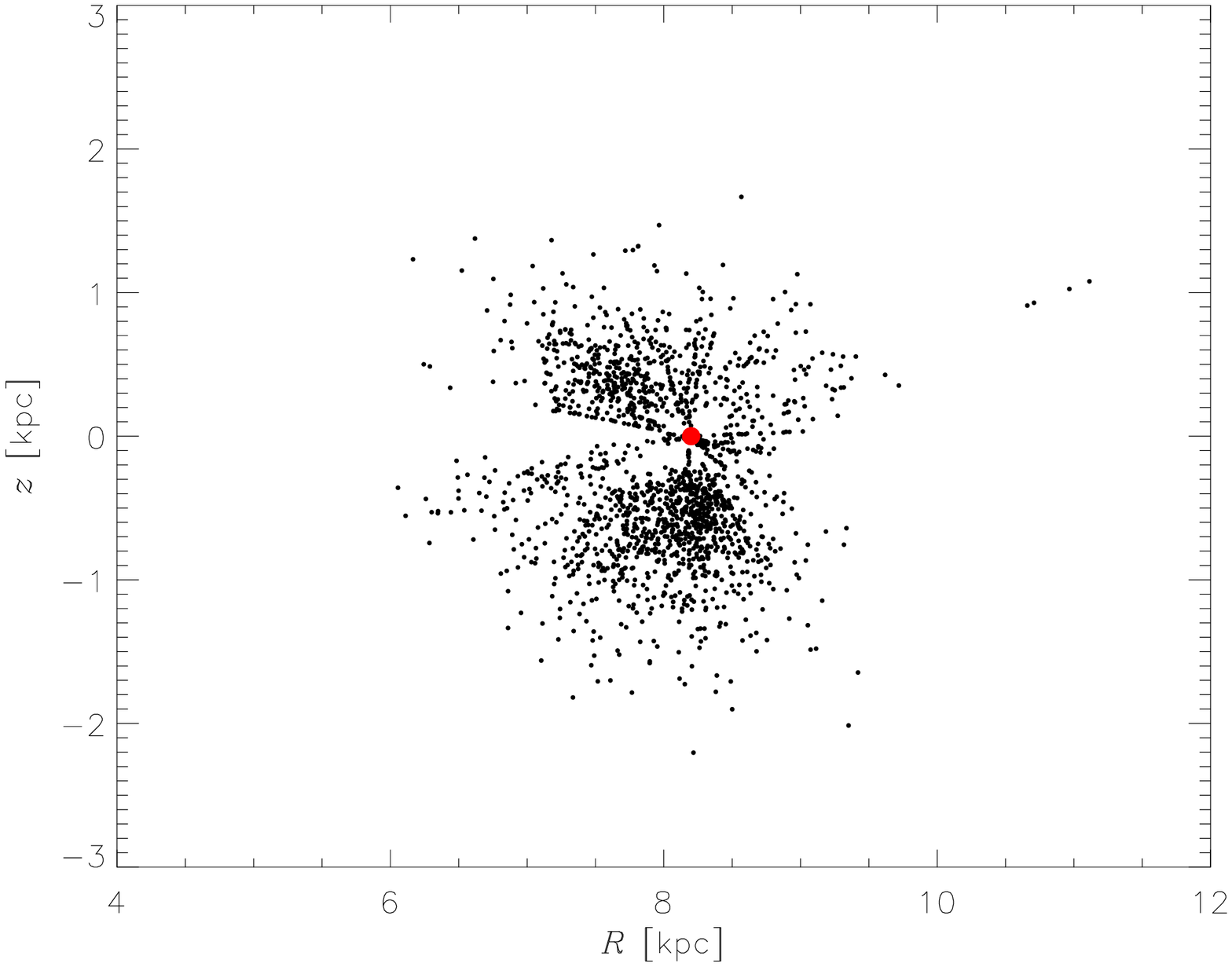}
\caption{Galactic radii and vertical distances from the Galactic plane for the sample of 1804 stars with accurate  $\pi, \mu_{\alpha}$, and $\mu_{\delta}$;
the red filled circle shows the Sun position. \label{fig:R_Z}}
\end{figure}

Starting from the obtained $\pi$, $\mu_{\alpha}$,  $\mu_{\delta}$, and $V_{\rm r}$ values, we computed the Galactic radii, $R$, the distance from the Galactic plane, $z$, and the  velocities $U_{\rm LSR}$, $V_{\rm LSR}$, and $W_{\rm LSR}$ together with their uncertainties, $\Delta_U$, $\Delta_V$, $\Delta_W$ using  a program kindly provided by Re Fiorentin (private communication). The program  assumes that the Sun is 8.2\,kpc
away from the MW centre, the LSR is rotating at 232\,km\,s$^{-1}$ around the Galactic centre \citep{McM17a, McM17b}, and the LSR peculiar velocity components of
the Sun are: ($U_{\Sun}$, $V_{\Sun}$, $W_{\Sun}$) = (-11.1, 12.24, 7.25) km\,s$^{-1}$ \citep{SCH10} in a right-handed coordinate system.
Figure\,\ref{fig:R_Z} shows the Galactic positions of our stars; most of the stars have  $R$ between 6.5 and 8.5\,kpc and $z$ between -2.0 and 1.5\,kpc.

From  $U_{\rm LSR}$, $V_{\rm LSR}$, and $W_{\rm LSR}$ and their uncertainties we  computed, using equations\,\ref{eq:fProb} and \ref{eq:Prob}, the probability $P_{\rm Thin}$, $P_{\rm Thick}$, $P_{\rm Halo}$, $P_{\rm Hercules}$ 
and their uncertainties $\sigma_{P_{\rm Thin}}$,  $\sigma_{P_{\rm Thick}}$,  $\sigma_{P_{\rm Halo}}$, $\sigma_{P_{\rm Hercules}}$.
Both in computing $\Delta_U$, $\Delta_V$, $\Delta_W$ and  $\sigma_{P_{\rm Thin}}$,  $\sigma_{P_{\rm Thick}}$,  $\sigma_{P_{\rm Halo}}$, $\sigma_{P_{\rm Hercules}}$
we did not applied the standard error propagation but we used the proper co-variance matrices due to the fact that the variables are correlated.

We then define a sample of 1356 stars as representative  of the thin disk population (hereafter, called Thin$^{\rm K}$ sample, where the superscript ``K'' recall that it was obtained via a Kinematical  selection) by using the following criteria:

 \begin{eqnarray} \label{eq:Thin}
& P_{\rm Thin}-\sigma_{P_{\rm Thin}}& >  2\,(P_{\rm Thick} +\sigma_{P_{\rm Thick}}) \nonumber\\
& ~~~~~~~~~~~~~~~~~~~~~~ & >  2\,(P_{\rm Halo} +\sigma_{P_{\rm Halo}})  \\
& ~~~~~~~~~~~~~~~~~~~~~~ & >  2\,(P_{\rm Hercules} +\sigma_{P_{\rm Hercules}}) \nonumber
  \end{eqnarray}

In analogy we define a sample of 196 stars as representative of the thick disk population (Thick$^{\rm K}$ sample) by using the following criteria:
 
 \begin{eqnarray} \label{eq:Thick}
& P_{\rm Thick}-\sigma_{P_{\rm Thick}} & >  2\,(P_{\rm Thin} +\sigma_{P_{\rm Thin}})  \nonumber\\
& ~~~~~~~~~~~~~~~~~~~~~~~~ & >  2\,(P_{\rm Halo} +\sigma_{P_{\rm Halo}}) \\
& ~~~~~~~~~~~~~~~~~~~~~~~~ & >  2\,(P_{\rm Hercules} +\sigma_{P_{\rm Hercules}})  \nonumber
  \end{eqnarray} 

Adopting a similar criterion we found that 6 stars of our sample belong to the Halo and 4 stars belong to the Hercules stream. The factor 2 in equations\,\ref{eq:Thin} and \ref{eq:Thick} is introduced to minimize the contamination between the two samples. It must be noticed that our selection criteria are more robust than those in general adopted by other authors, like for example \citet{BEN14}, since we take into account also the uncertainties on the $P$ probabilities due to the $\Delta_U$, $\Delta_V$, $\Delta_W$ values. The average  3D-NLTE corrections for the Thin$^{\rm K}$  and Thick$^{\rm K}$ samples are $-0.02\pm 0.02$\,dex and $-0.03\pm 0.01$\,dex, respectively.

\subsection{Samples selection on the bases of orbital parameters}
\label{sec:Orbits}
In the previous section we discriminated between thin and thick disk stars by using the present stellar Galactic velocities. Obviously, the stars during their life change position and velocity due to their motion in the Galaxy. In order to take into account this fact we compute stellar Galactic orbits for the 1804 stars with known distances and Galactic velocities. Orbital parameters for each star (maximum and minimum galactocentric radii, $R_{\rm max}$ and $R_{\rm min}$, maximum absolute distance from  Galactic plane, $|Z_{\rm max}|$, and orbital
eccentricity, $\epsilon$) were calculated using a code kindly provided us by J.\,P.\,Fulbright (private communication). The code uses an integrator developed  by D. Lin firstly used in \citet{FUL02} and assumes a three-component potential (halo, disk, and bulge) based on the potential described in \citet{JOH96} and \citet{JOH98}.
Each star is followed for 15\,Gyr, at a step of 3\,Myr. It is well known that stellar migration, via churning and blurring, make difficult to estimate the birth  radius of each star and therefore its identification with $R_{\rm med}=0.5\times(R_{\rm min}+R_{\rm max}$) is not straightforward. In particular, we do not have in our Galactic potential deviations from axisymmetry like those introduced by the bar and the spiral arms. Therefore, we will use in the following  only  $|Z_{\rm max}|$ and $\epsilon$ which we assume, in a first approximation, not significantly affected by stellar migration.

\begin{figure}[htbp]
\includegraphics[width=1\textwidth]{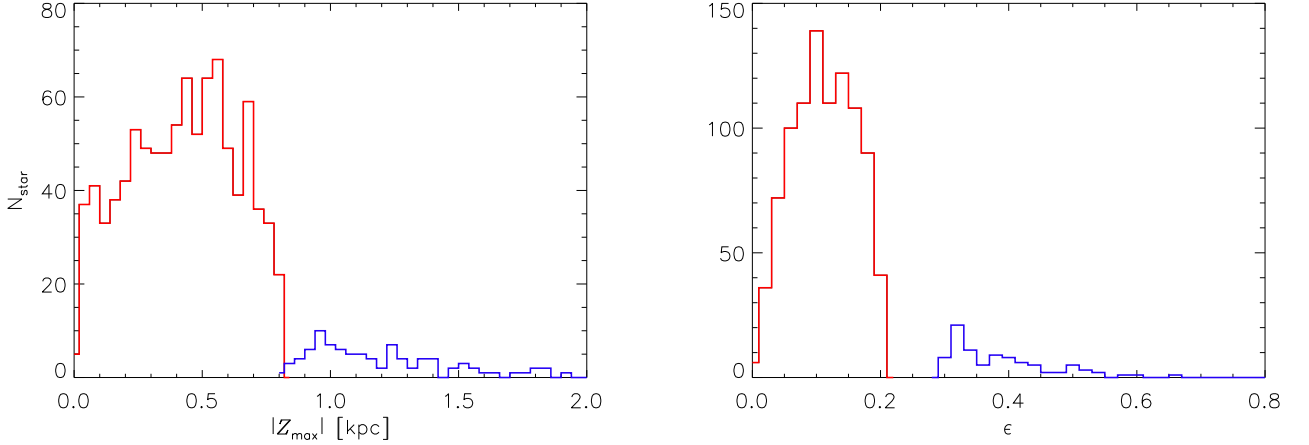}
\caption{Distributions of $|Z_{\rm max}|$ and eccentricity, $\epsilon$, for Thin$^{\rm O}$ (red) and  Thick$^{\rm O}$ (blue) samples. \label{fig:Zmax_ecc}}
\end{figure}

\begin{figure}[htbp]
\includegraphics[width=1\textwidth]{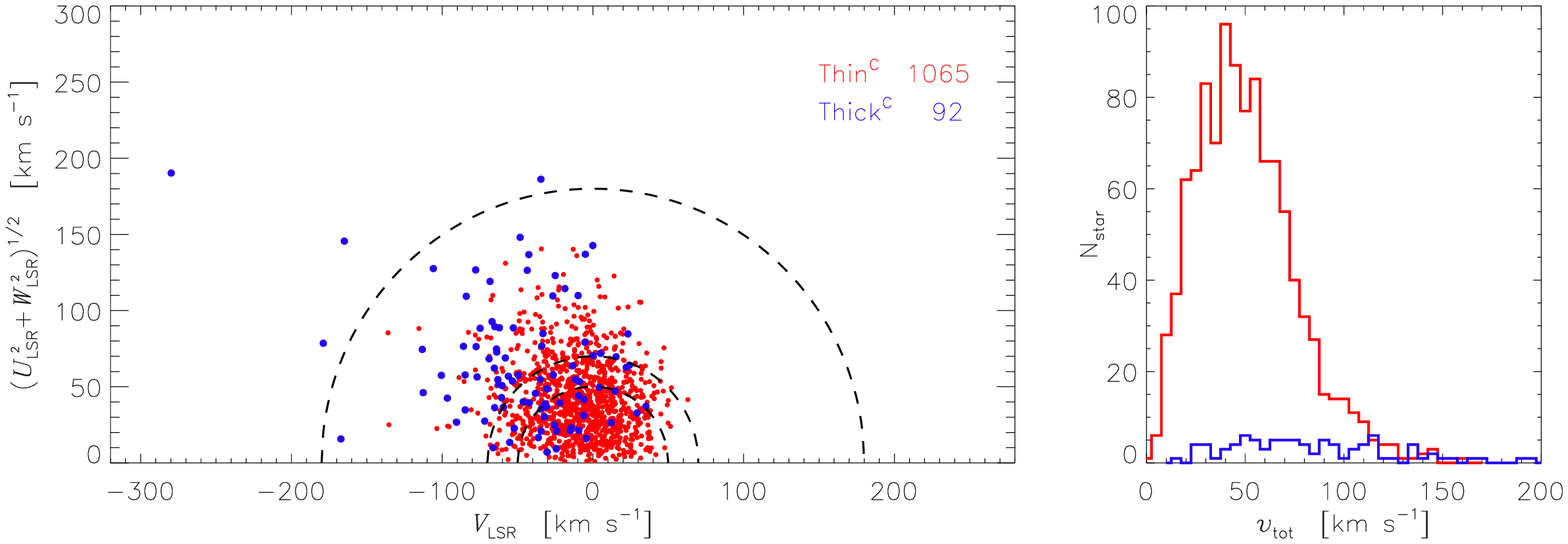}
\includegraphics[width=1\textwidth]{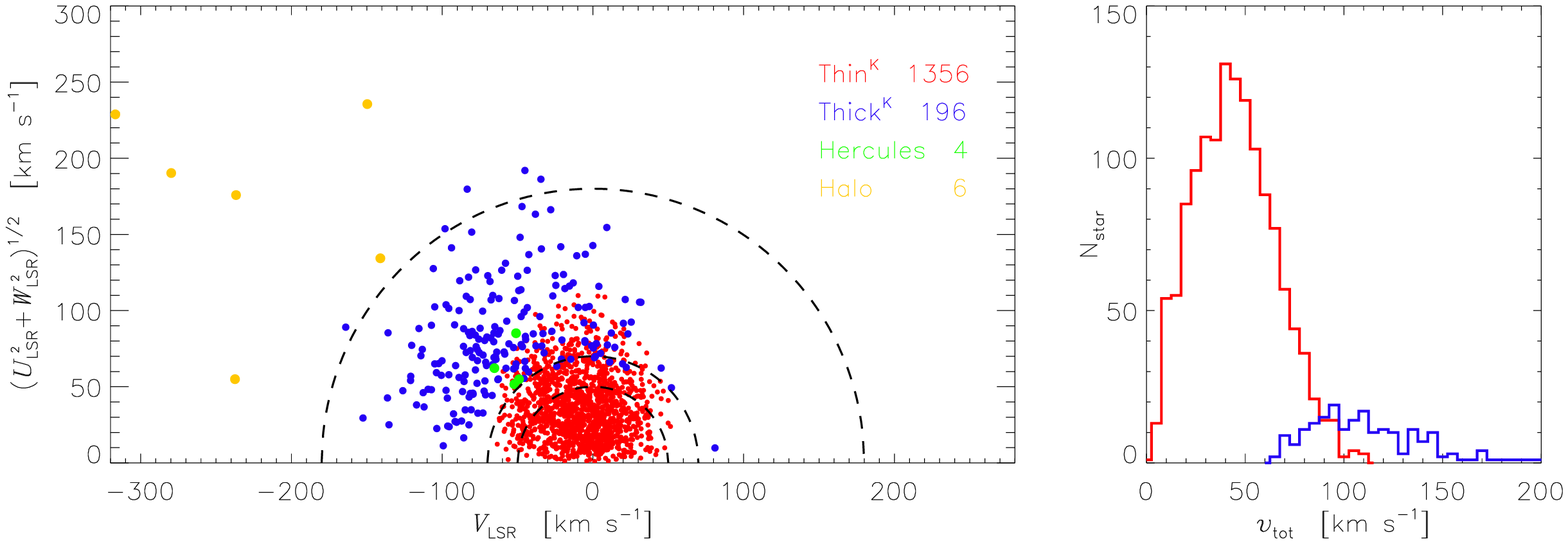}
\includegraphics[width=1\textwidth]{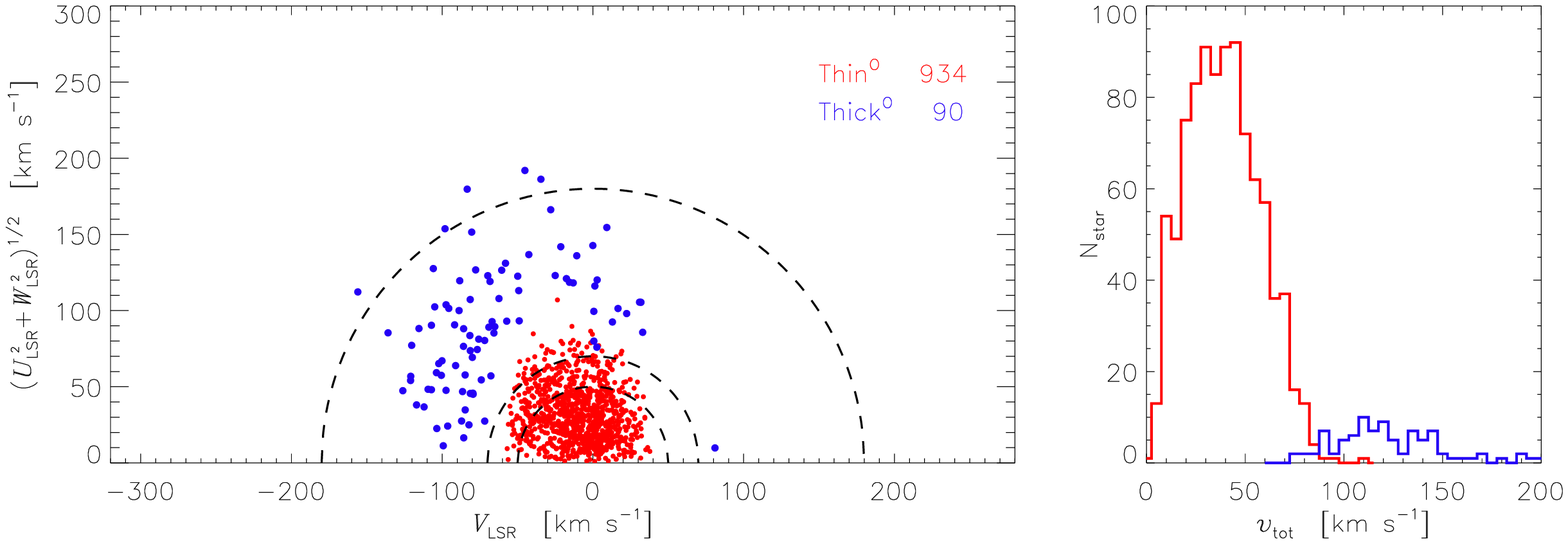}
\caption{Toomre diagrams for the the thin (red points) and thick (blue points) disk stars with different selection. Dotted lines show constant values of the total space velocity,  
$v_{\rm tot}$ $\equiv$ ($U_{\rm LSR}^2+V_{\rm LSR}^2+W_{\rm LSR}^2$)$^{1/2}$ at 50, 70 and 180 km s$^{-1}$ i.e. the thresholds used by \citealt{NIS04} to separate as a first approximation, thin and thick disk stars (see Section\,\ref{sec:kinematics}).
Upper panels: vertical and radial kinetic energy versus rotational ones (left panel) and distribution of total velocities for stars belonging to  Thin$^{\rm C}$ and  Thick$^{\rm C}$ samples with available Galactic velocities.  Middle panels: the same  as the upper panels but for Thin$^{\rm K}$ and  Thick$^{\rm K}$ samples; stars belonging to Hercules stream (green points) and halo (yellow points) are also indicated in the left panel. Lower panels:  the same  as the upper panels but for  Thin$^{\rm O}$ and  Thick$^{\rm O}$ samples.  \label{fig:Toomre}}
\end{figure}

To select the Thin$^{\rm O}$ and  Thick$^{\rm O}$ samples we used the plane $\epsilon$ vs $|Z_{\rm max}|$. We fix an upper limit for the 
$|Z_{\rm max}|$ of the thin disk stars at 0.80\,kpc which is the value where the stellar densities of the two disks, computed using the thin- and thick-disk model of \citet{WID12} with scale heights H$_1$ and H$_2$ from \citet{FER17}, are equal. Then, to remove the contamination of thick disk stars with low $|Z_{\rm max}|$, we require also an eccentricity lower than 0.2 (see discussion in \citet{WIL11}). For the thick disk star selection, we require $0.8<|Z_{\rm max}|< 2.0$\,kpc and $0.3<\epsilon<0.7$ where the upper values are needed to exclude any halo star. With such criteria we obtain two samples,  Thin$^{\rm O}$ and  Thick$^{\rm O}$ (see Figure\,\ref{fig:Zmax_ecc}), containing 934 and 90 thin and thick disk stars with average  3D-NLTE corrections of $-0.02\pm 0.02$\,dex and $-0.03\pm 0.01$\,dex, respectively. 

The comparison of the kinematical and chemical properties of the three pairs of Thin and Thick samples is presented in the following Sections while the percentages of stars in common between the different selections is given in Section\,\ref{sec:conclusion}.

\section{Kinematical properties of the thin and thick disk star samples } \label{sec:kinematics}
To study the kinematical properties of the different Galactic populations, a commonly used tool
is the Toomre diagram, which is a representation of 
the combined vertical and radial kinetic energies versus the rotational energy. As a first approximation,   the low-velocity stars, with a total velocity $v_{\rm tot}$ $\equiv$ ($U_{\rm LSR}^2+V_{\rm LSR}^2+W_{\rm LSR}^2$)$^{1/2}$ less than 50 km s$^{-1}$, are mainly thin disk stars while the stars with 70 $\leq$  $v_{\rm tot}$ $\leq $ 180 km\,s$^{-1}$ are likely to be thick disk stars (e.g. \citealt{NIS04}). 
Moreover,  the thick disk is as a whole a more slowly rotating stellar system than the thin disk lagging behind the LSR by approximately 
50\,km\,s$^{-1}$ (e.g. \citealt{SOU03}).
The Galactic velocity dispersions ($\sigma_U$, $\sigma_V$, $\sigma_W$) are also larger in the thick disk than in the thin
disk. For example, \citet{SOU03} found ($\sigma_U$, $\sigma_V$, $\sigma_W$ = 39 $\pm$ 2, 20 $\pm$ 2, 20$\pm$1\,km\,s$^{-1}$)  and 
($\sigma_U$, $\sigma_V$, $\sigma_W$  =  63 $\pm$ 6, 39 $\pm$ 4, 39 $\pm$ 4\,km\,s$^{-1}$)  for the thin and thick disks respectively,  in reasonable agreement with the values by \citet{BEN14}
($\sigma_U$, $\sigma_V$ , $\sigma_W$  =  35, 20, 16\,km\,s$^{-1}$), and ($\sigma_U$, $\sigma_V$, $\sigma_W$ = 67, 38, 35\,km\,s$^{-1}$) used in Section\,\ref{sec:Gal_Vel} . 

\begin{deluxetable}{lcccr}
\tablecaption{Dispersion velocities. \label{tab:disper_vel}}
\tablecolumns{5}
\tablewidth{0pt}
\tablehead{
\colhead{} &
\colhead{$\sigma_U$} &
\colhead{$\sigma_V$} &
\colhead{$\sigma_W$} &
\colhead{~~N$_{\rm star}$}\\[-8pt]
\colhead{} &
\colhead{~km\,s$^{-1}$~} &
\colhead{~km\,s$^{-1}$~} &
\colhead{~km\,s$^{-1}$~} &
\colhead{}
}
\startdata
&&&&\\[-8pt]
  ~~~~~Thin$^{\rm C}$~~~~~ & 42   & 27 &    21 & 1065 \\
 ~~~~~Thin$^{\rm K}$~~~~~ &         38    &    23   &     17 & 1356 \\
~~~~~Thin$^{\rm O}$~~~~~ &         31    &    20   &     18 & 934 \\
~~~~~Thin$^{\rm S}$~~~~~ &        39     &   20    &    20 &\\
~~~~~Thin$^{\rm B}$~~~~~ &     35 & 20 & 16 &\\[8pt]
  ~~~~~Thick$^{\rm C}$~~~~~ & 64 & 48 & 41 & 92 \\
 ~~~~~Thick$^{\rm K}$~~~~~ &         68   &   42   &   51 & 196\\
 ~~~~~Thick$^{\rm O}$~~~~~ &         86   &   45  &   41 & 90 \\
 ~~~~~Thick$^{\rm S}$~~~~~ &        63  &  39 &   39 & \\
 ~~~~~Thick$^{\rm B}$~~~~~ & 67 & 38 & 35 &  \\[2pt]
\enddata
\tablecomments{Thin$^{\rm S}$ and Thick$^{\rm S}$ from \citet{SOU03}; Thin$^{\rm B}$ and Thick$^{\rm B}$ from \citet{BEN14}.}
\end{deluxetable}

In the three left panels of Figure\,\ref{fig:Toomre} we show the positions in the Toomre diagram 
of the stars we attribute to thin or thick disk by using the different selection criteria described in Sections\,\ref{sec:Adibek}, \ref{sec:Gal_Vel}, and \ref{sec:Orbits}.   As can be seen, stars belonging to the thin disk samples and those belonging to 
the thick disk samples are fairly kinematically  separated. The separation is less clear in the upper panel which refers to the chemical selection. It is worthwhile noticing that this selection does not take into account any kinematical stellar property. Note, also, that  we do not have Gaia DR2 data for all  the Thin$^{\rm C}$ and  Thick$^{\rm C}$ stars and thus the corresponding Toomre diagram and  histograms contain only 1065 and 92 thin and thick disk stars, respectively instead of 1267 and 99. As far as the other two selections are concerned, while the separation in the middle panel is expected since the kinematical selection is actually based on the stellar Galactic velocities, the clear segregation in the lower panel is less predictable even if not completely unexpected. In the three selection cases there is always a common interval in the $v_{\rm tot}$ distributions (right panels)  of the two samples with a decreasing overlapping going from top to bottom. 
The widest overlap is obtained in a region $0<v_{\rm tot} \lesssim 130$\,km s$^{-1}$ when using the chemical selection; the overlap region  is 70$ \lesssim v_{\rm tot} \lesssim 100$\,km s$^{-1}$ for the kinematical selection and reduces to $70 \lesssim v_{\rm tot} \lesssim 90$\,km\,s$^{-1}$ for orbital selection. It is also worthwhile mentioning that
 the difference between the mean rotational velocity of the two samples ($<V_{\rm thin} >$ - $<V _{\rm thick}>$) is 36, 50, and 55\,km\,s$^{-1}$ for the chemical, kinematical and orbital selection, respectively. In Table\,\ref{tab:disper_vel} we report the dispersion velocities of the different samples, ($\sigma_U^{\rm C}$, $\sigma_V^{\rm C}$, $\sigma_W^{\rm C}$), ($\sigma_U^{\rm K}$, $\sigma_V^{\rm K}$, $\sigma_W^{\rm K}$), and  ($\sigma_U^{\rm O}$, $\sigma_V^{\rm O}$,  $\sigma_W^{\rm O}$); independently of the adopted selection the thick disk stars show always larger dispersion velocities than the thin disk ones as expected (see for example \citet{SOU03} and \citet{BEN14}). 

\section{Chemical properties of the thin and thick disk star samples} \label{sec:C_result}

\begin{figure}[htbp]
\includegraphics[width=1\textwidth]{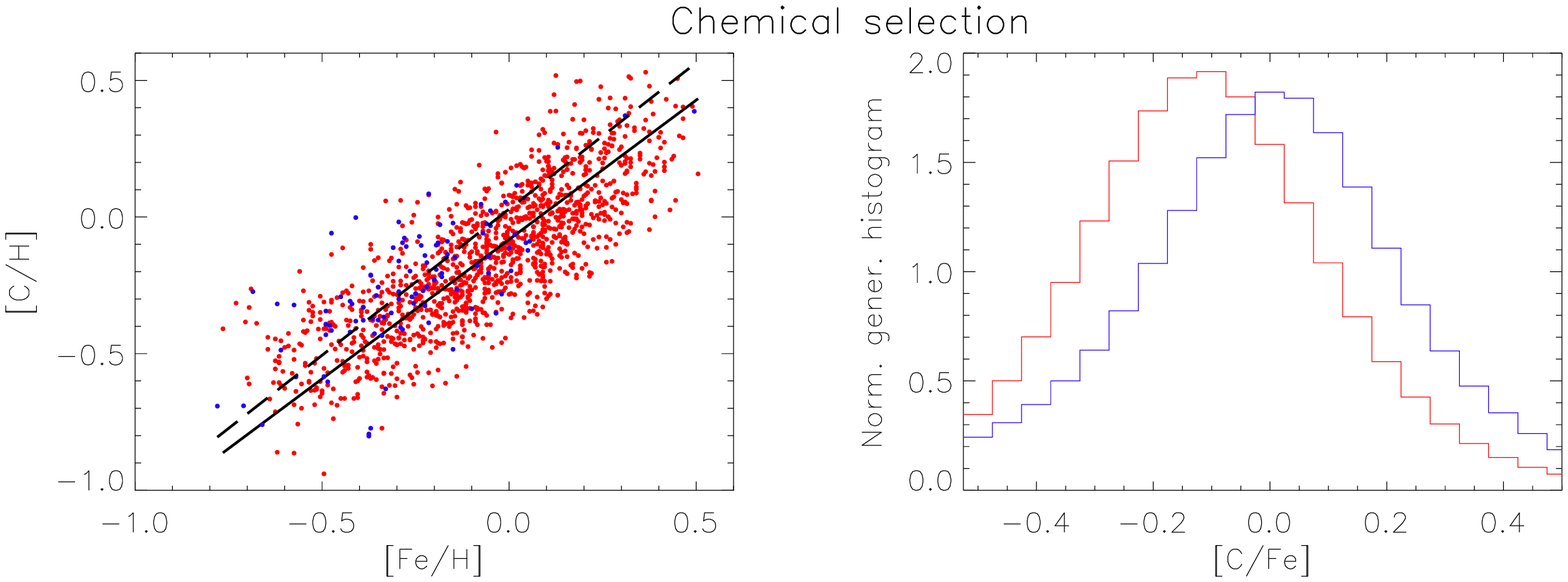}
\includegraphics[width=1\textwidth]{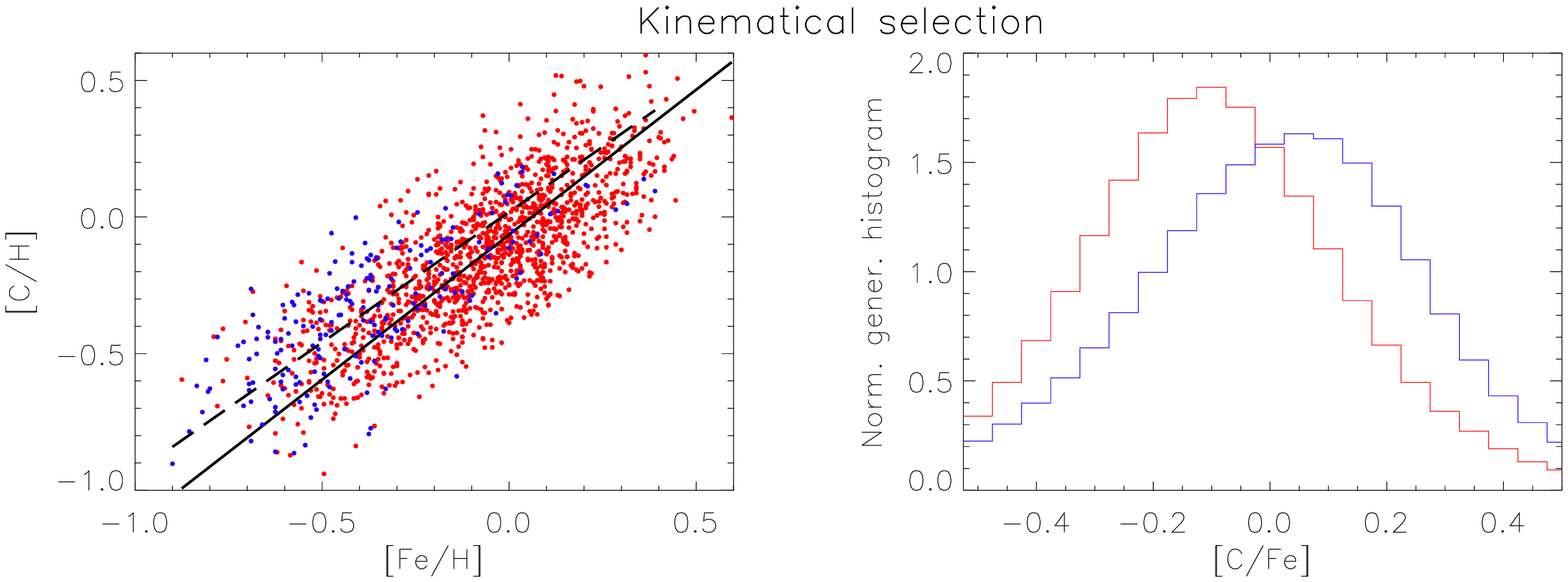}
\includegraphics[width=1\textwidth]{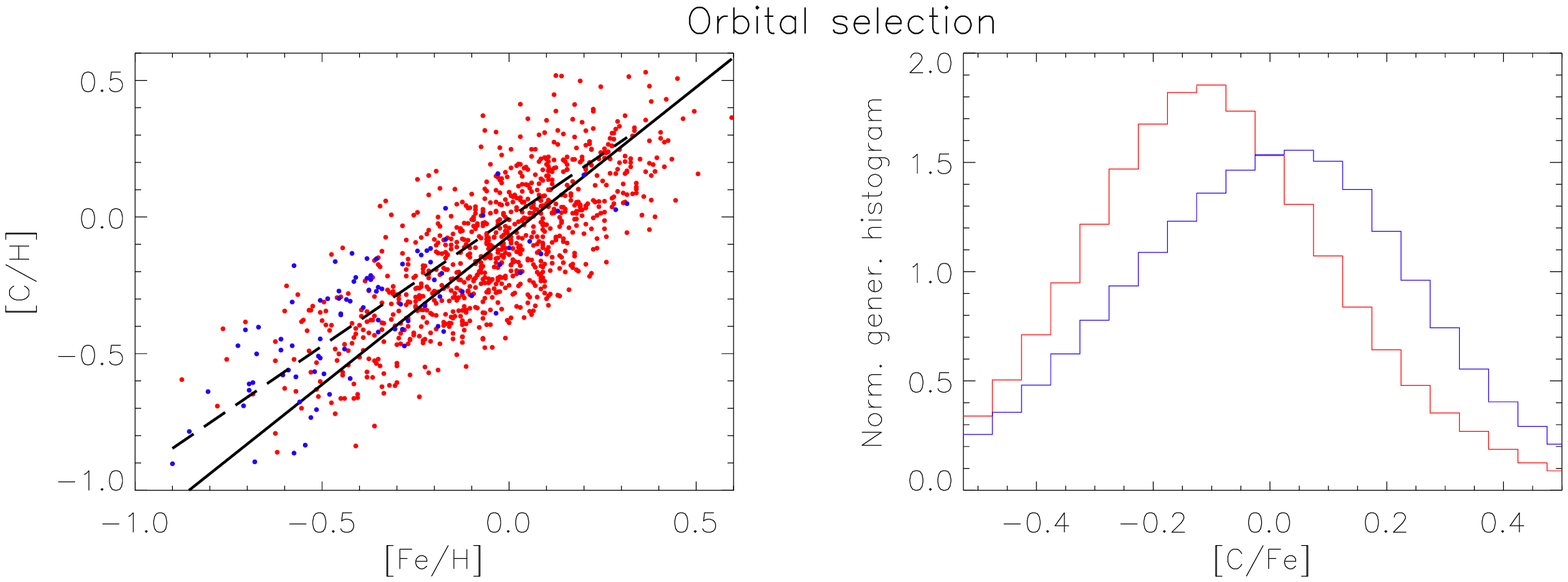}
\caption{[C/H]--[Fe/H] diagrams  (left panels) and  [C/Fe] normalized generalized histograms (right panels) for  thin (red) and thick (blue) disk samples.  
Upper panels:   Thin$^{\rm C}$ and  Thick$^{\rm C}$ samples;  middle panels:  Thin$^{\rm K}$ and  Thick$^{\rm K}$ samples; lower panels: Thin$^{\rm O}$ and  Thick$^{\rm O}$ samples. Regression lines for thin (continuous)  and thick (dashed) samples are superimposed on the left panels.
\label{fig:CH}}
\end{figure}

The left panels of Figure\,\ref{fig:CH} show the [C/H] vs [Fe/H] for the thin and thick disk samples. As can be seen, and evidenced by the regression lines,  thick disk stars have larger C abundance than thin disk stars at the same [Fe/H] on average. The regression lines have slopes which differ by less than 2-sigma, therefore, the slope differences may be not significant. In the right panels we show the normalized generalized distributions of [C/Fe] built  by summing individual unit area Gaussian computed for each star, in the proper sample, by using its [C/Fe] value and uncertainty and, then, normalizing the results to the number of objects. The thin disk sample (red) distribution is, for any kind of selection, peaked at lower [C/Fe] than the thick disk sample distribution (blue). A two--sided Kolmogorov-Smirnov test, performed using the ``kstwo'' IDL\footnote{Interactive Data Language: https://www.harrisgeospatial.com/Software-Technology/IDL} routine, confirms that the two cumulative distribution functions are significantly different for all the three different selections methodologies adopted (prob always less than 1.E-8).  
The left panels show for both the thin and thick disk stars a large scatter in the [C/H] values. Such a scatter was also found by \citet{NIS18,AMA19b} and they suggested that it can be explained by  variations in the dust to gas ratio in different star--forming gas clouds and/or by the need of applying differential 3D non--LTE corrections to 1D LTE abundances (see discussion at the end of Section\,\ref{sec:CFe_det}).

\begin{figure}[htbp]
\includegraphics[width=0.5\textwidth]{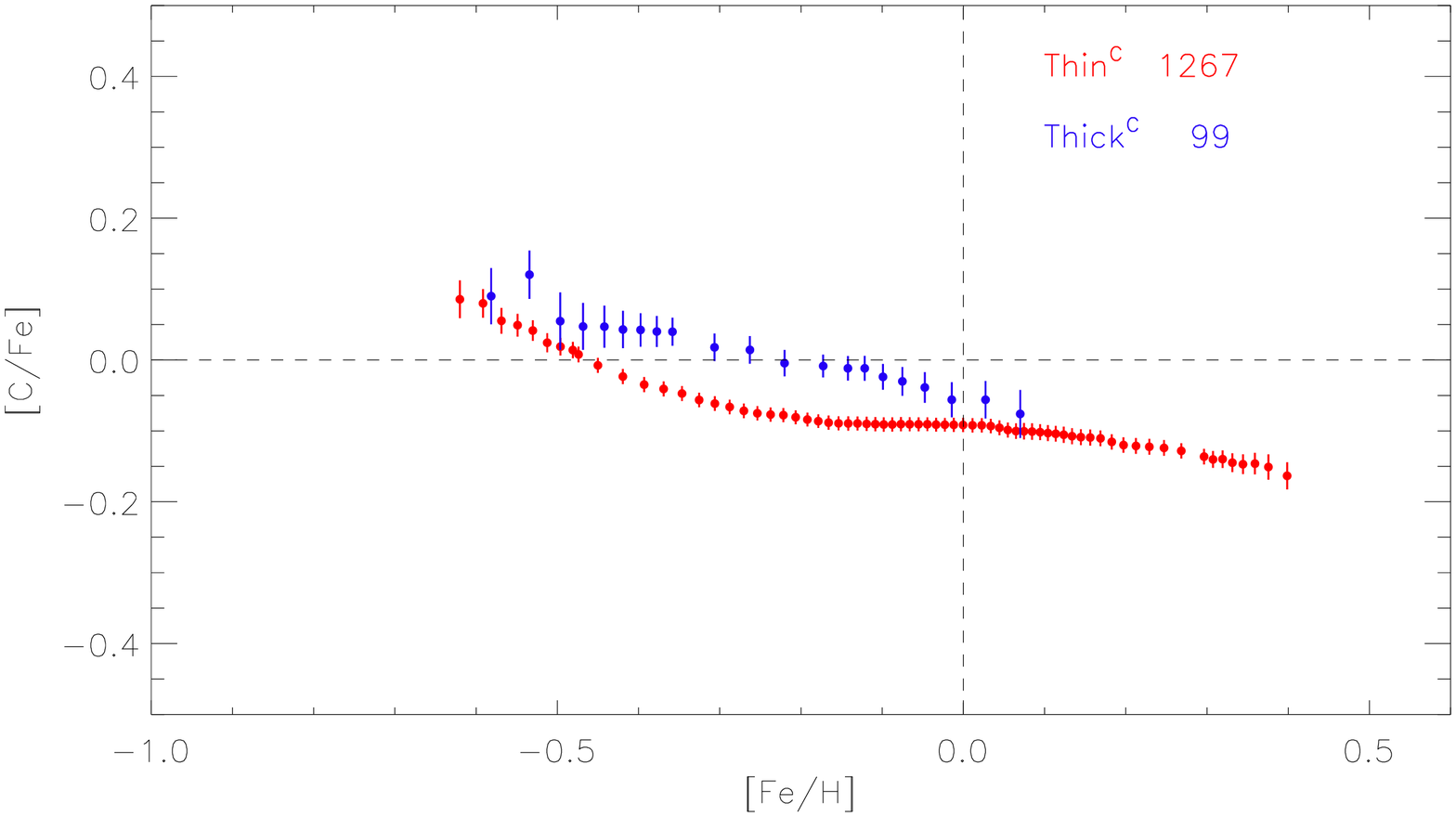}
\includegraphics[width=0.5\textwidth]{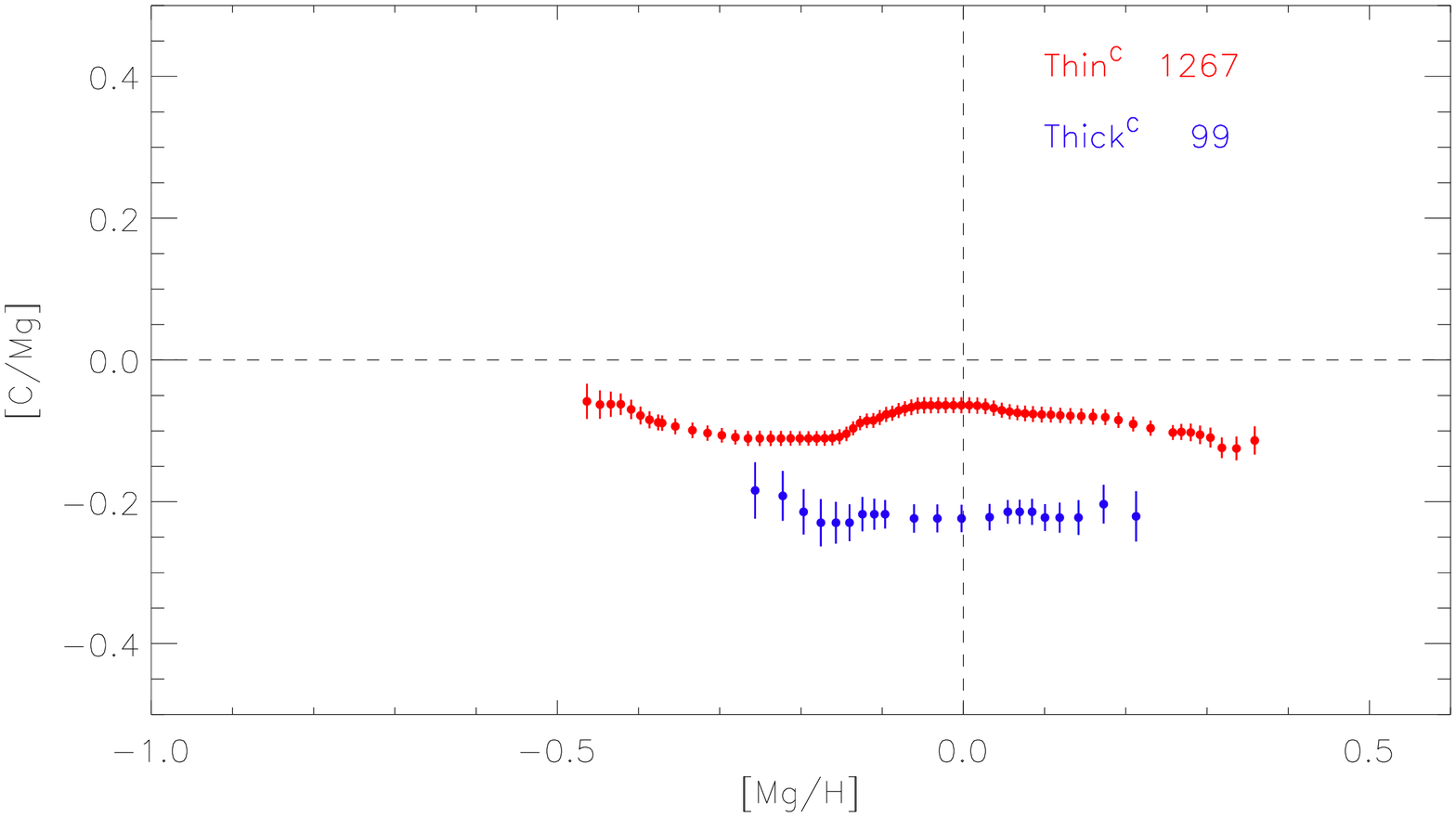}
\includegraphics[width=0.5\textwidth]{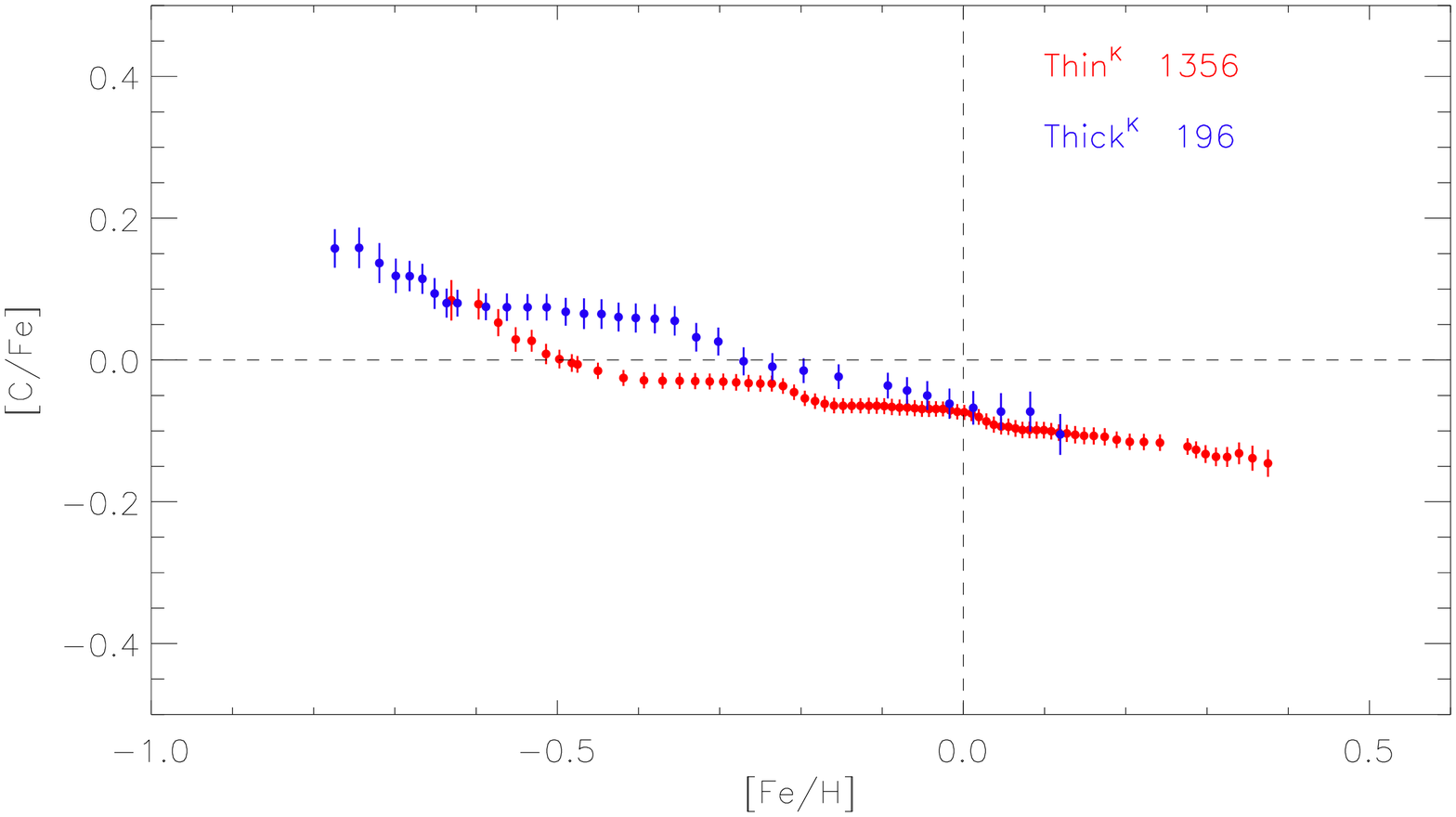}
\includegraphics[width=0.5\textwidth]{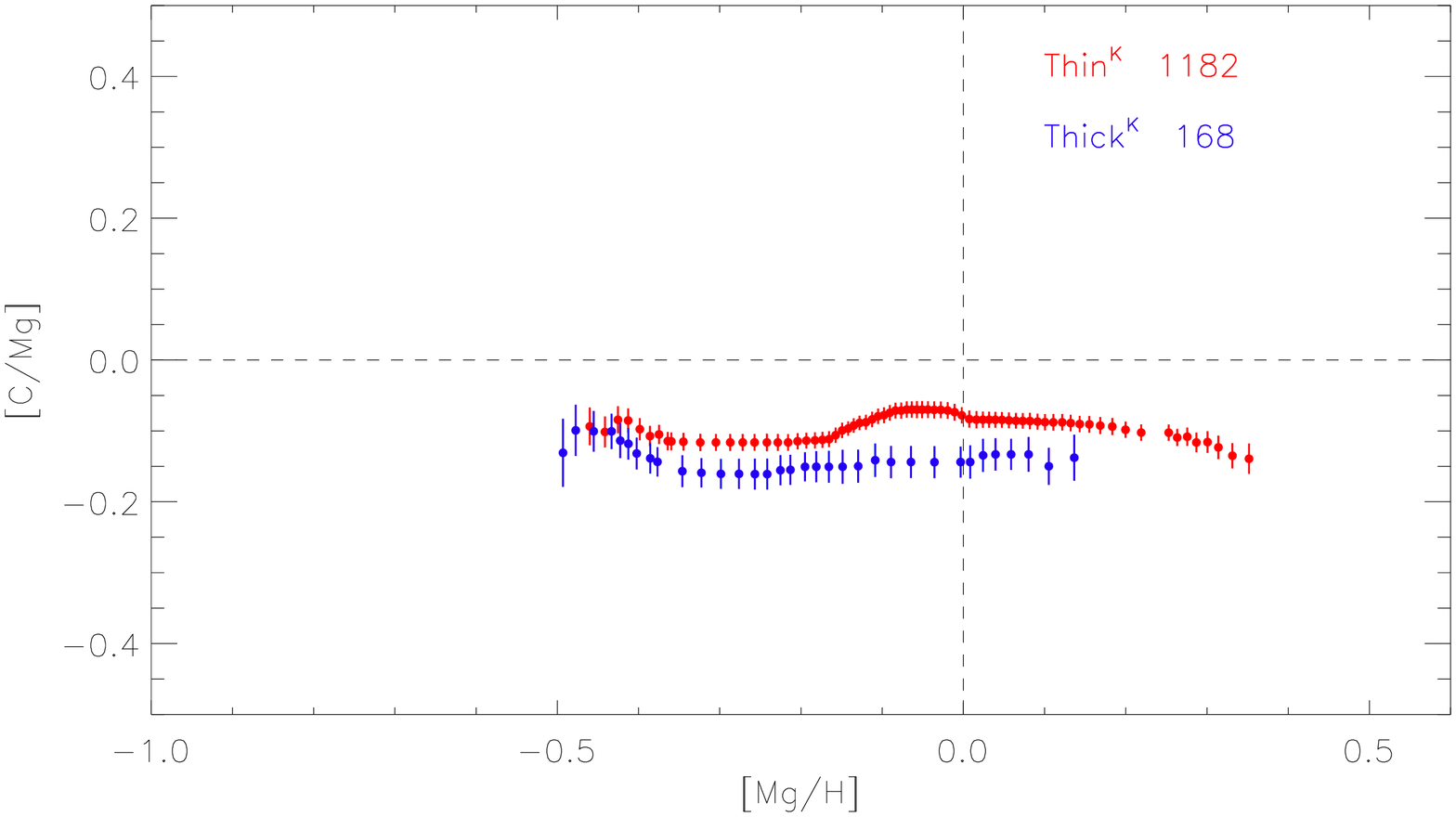}
\includegraphics[width=0.5\textwidth]{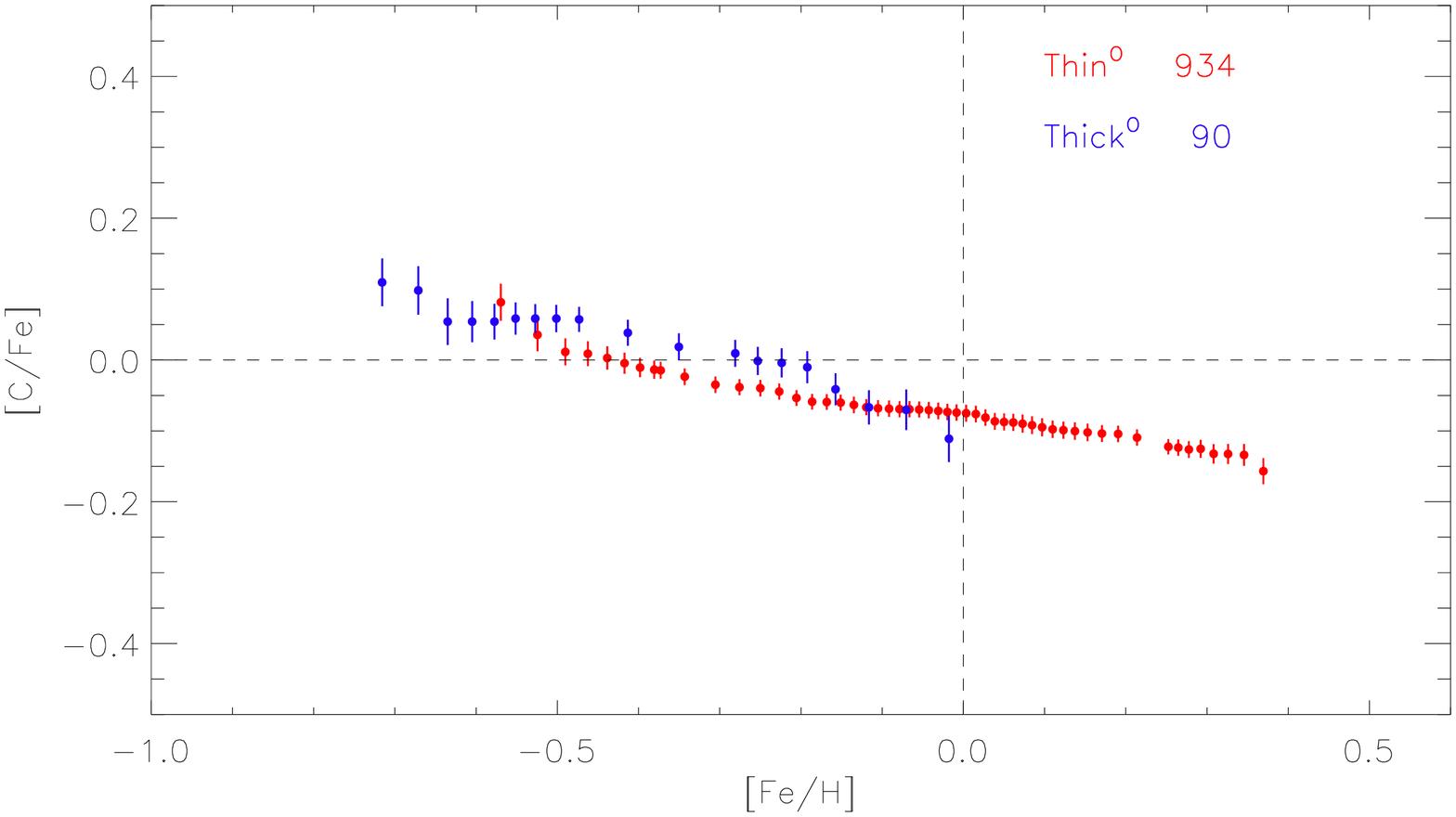}
\includegraphics[width=0.5\textwidth]{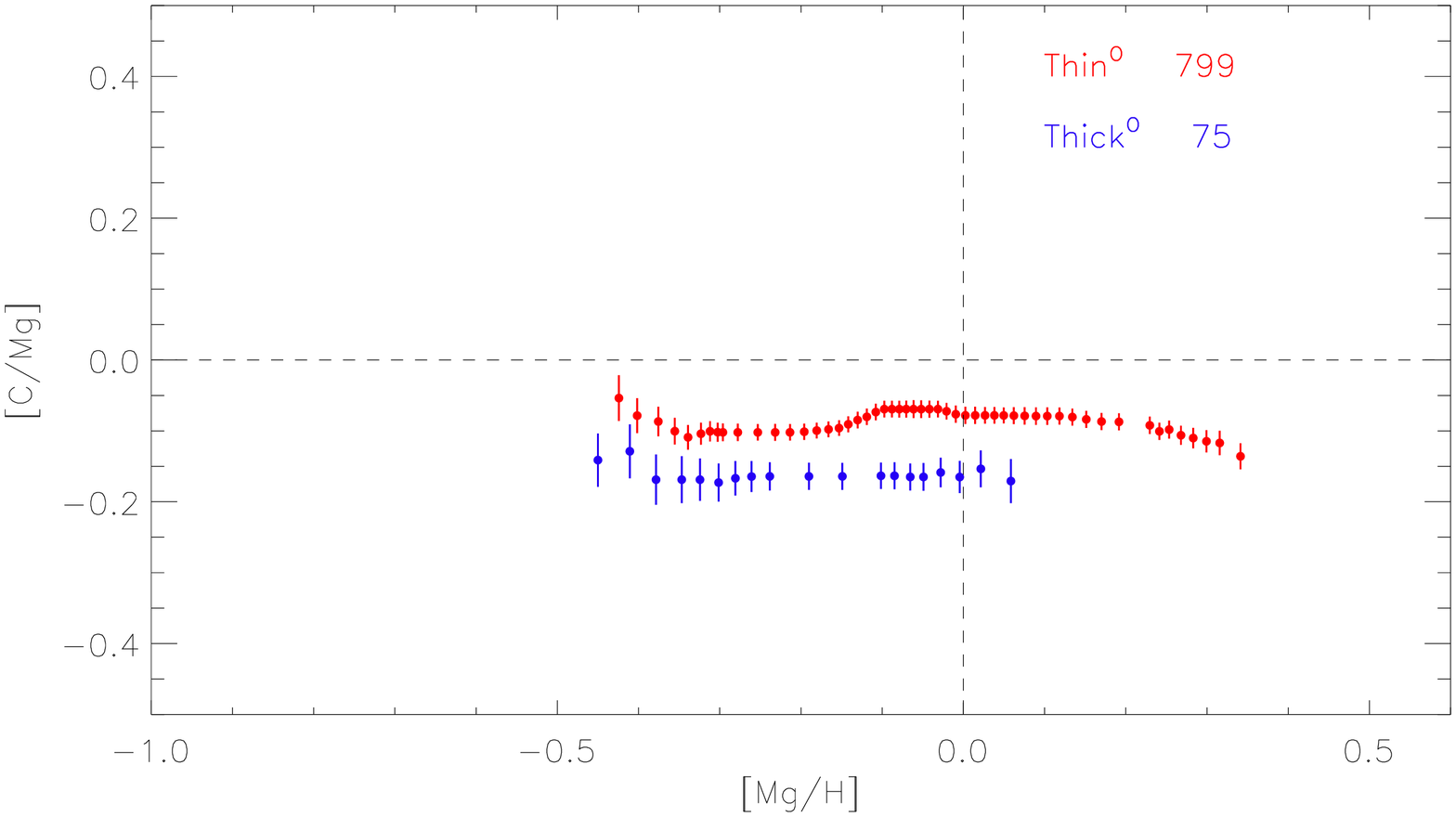}
\caption{[C/Fe]--[Fe/H] diagrams  (left panels) and   [C/Mg]--[Mg/H] diagrams  (right panels) for  thin (red) and thick (blue) samples: binned running averages and standard deviations are plotted.
Upper panels:   Thin$^{\rm C}$ and  Thick$^{\rm C}$ samples;  middle panels:  Thin$^{\rm K}$ and  Thick$^{\rm K}$ samples; lower panels:  Thin$^{\rm O}$ and  Thick$^{\rm O}$ samples.
\label{fig:CMg_Mg}}
\end{figure}

To better understand the behaviour of [C/Fe] we plotted in Figure\,\ref{fig:CMg_Mg} the trends of [C/Fe] vs [Fe/H] for the thin and thick samples for the three different selection. In order to get rid of the quite large scatter in the data shown in Figure\,\ref{fig:CH}, we opted to plot, instead of the  individual values, the mean ones in partially overlapped bins  by using a running average (using a fixed number of points) together with, for each bin, their standard deviations. As can be seen, for all the selections, the thick disk stars show a higher [C/Fe] than the thin disk stars for $-0.5\lesssim$\,[Fe/H]\,$\lesssim -0.1$. In the case of the chemical selection the [Fe/H] range, for which the thick and thin trends are separated, extends to [Fe/H]$\simeq +0.1$ while, in the case of the kinematical selection, we have an intermediate situation between the other two cases. The so obtained [C/Fe] vs [Fe/H] trends recall the behaviour of $\alpha$--elements vs [Fe/H] but with a less pronounced separation.  It is worth noticing that in all the panels the thin disk sequences at [Fe/H]=0 fall below the zero horizontal line. This offset, was also found in other literature works (e.g.  \citet{SHI02}, \citet{NIS14}, and \citet{NIS18}), suggests that, maybe, the Sun can be carbon rich with respect to the average thin disk \citep[see also Fig.\,2 in][]{MEL09}.

Our results are in agreement with those by \citet{RED06}, \citet{DEL10}, \citet{NIS14}, and \citet{SUA17}. In particular, on the basis of a  kinematical selection, \citet{RED06} found that the abundance ratios [C/Fe] for their thick disk sample stars with [Fe/H] $< -0.4$ were, on average, larger than for their thin-disk stars of the same [Fe/H]. They also stated that carbon behaves like Mg and other $\alpha$-elements.  \citet{NIS14}, by also implementing a kinematical selection for some stars and a chemical one for other stars,  found that their thin disk stars fall below the thick disk ones in the [C/Fe] vs [Fe/H] diagram for [Fe/H]$\sim -0.3$ and suggested that the two populations merge at higher metallicities. On the other hand, our results contradict those by \citet{BEN06} who, by using a kinematical selection, found that the [C/Fe] vs [Fe/H] trends for the thin and thick disks are totally merged and flat for sub--solar metallicities with a shallow decline for the thin disk stars from [Fe/H]$\simeq 0$  up to [Fe/H] $\simeq +0.4$. On the contrary our data show a general decrease of [C/Fe] for both thin and thick disk stars with increasing [Fe/H] values even if the upper- and middle--left panels seem to indicate a flattening for the thin disk stars with -0.5$\lesssim$\,[Fe/H]\,$\lesssim 0$.

In order to study the origin and Galactic evolution of carbon, the [C/O]-[O/H] diagram is often used in the literature. In fact, since oxygen is exclusively produced in massive stars on a relatively short timescale, the change in [C/O] as a function of [O/H] gives hints on the yields and timescales of carbon production in different types of stars \citep[see][and references therein]{CES09}. Nevertheless, the derivation of oxygen abundances for dwarf stars within GES iDR5 has not yet been completed, \citep[for giants see][]{MAG18}, therefore we decided to use Magnesium instead of oxygen. The right panels of Figure\,\ref{fig:CMg_Mg} show the trends of [C/Mg] vs [Mg/H] for the thin and thick samples for the three different selection.  As can be seen, for all the selections, the thin disk stars show an higher [C/Mg] than the thick disk stars for the same [Mg/H]. In the case of the chemical selection the difference in [C/Mg] is probably enhanced by the fact that  the thick and thin disk stars are actually high- and low-Mg stars. On the other hand the presence of a separation also in the other selection cases indicates that there is an excess of [C/Mg] in the thin disk stars with respect to the thick disk. 
All the panels show almost flat trends confirming the similarity of the C and $\alpha$--elements behaviours suggested by the left panels of Figure\,\ref{fig:CMg_Mg}.   

\section{Ages of the thin and thick disk stars samples}
\label{sec:age}
A better understanding of carbon evolution in our Galaxy could be achieved by determining the ages of the stars in our sample.  However, the ages of stars cannot be  directly measured and their determination, in particular for field stars,
is indeed very difficult \citep[see e.g.][and references therein]{RAN18}. 
A number of methods have been devised to derive stellar ages. Whilst  gyrochronology
and asteroseismology (or combinations of them) are recognized as the most reliable processes for dating
stars \citep{SOD10,ANG19}, such methods are not yet applicable to our stellar sample since
it is composed of field stars that still lack of oscillation data. We implement an isochrone comparison Bayesian approach, as firstly proposed by \citet{PON04} and \citet{JOR05}, of a
commonly used technique that is based on the comparison between observational quantities (e.g. magnitudes) 
and derived parameters (like effective temperatures) that has been extensively applied in the literature 
\citep[see e.g.][]{CAS11, HAY13, BEN17, HOW19}. 
On the basis of previous works it has been demonstrated that even small uncertainties on
$T_{\rm eff}$'s and magnitudes can result in large age errors, hence our approach aims at
deriving relative ages to provide valuable insights on the overall age characteristics of our stellar
samples. 

In this paper we use a program kindly provided us by L. Lindegren (private communication) based on the Bayesian age estimation code first described by \citet{JOR05}, assuming a flat metallicity prior due to the good precision of GES estimates \citep[see discussion in][]{JOR05}. The program, which uses Padova isochrones,  was modified by us in order to be able to use  as input data the Gaia G magnitude  by adopting the color--color transformations by \cite{EVA18}. 
The absolute G magnitudes of our stars, computed from Gaia DR2 parallaxes and  corrected for reddening by using the 3D Galactic extinction model by \citet{DRI03}, and  the GES iDR5 $T_{\rm eff}$'s were given in input to our program. 
In such a way we obtain  an age estimate, together with  the  full width half maximum of the its probability distribution (FWHM$_{\rm Age}$), for 1751 stars (53 stars have absolute magnitudes outside the range of our isochrone database). 
Then, to remove the most uncertain ages, we  discarded those stars (518) with FWHM$_{\rm Age}>8$\,Gyr which correspond to low main sequence stars where isochrones are overlapping thus affecting the accuracy of age determinations.  It is worth noticing that also the absolute individual ages of the remaining 1098 stars may still have quite large uncertainties due to systematic errors and inaccuracies in  the input data of the adopted Bayesian method and in the input physics of the isochrones. 
In order to cope with this problem we adopted the same technique used in Section\,\ref{sec:C_result} to get rid of the scatter, i.e. we computed mean ages and their standard deviations in bins  built by using a running average.

\begin{figure}[htbp]
\begin{center}
\begin{tabular}{ccc}
\includegraphics[angle=90,width=0.30\textwidth]{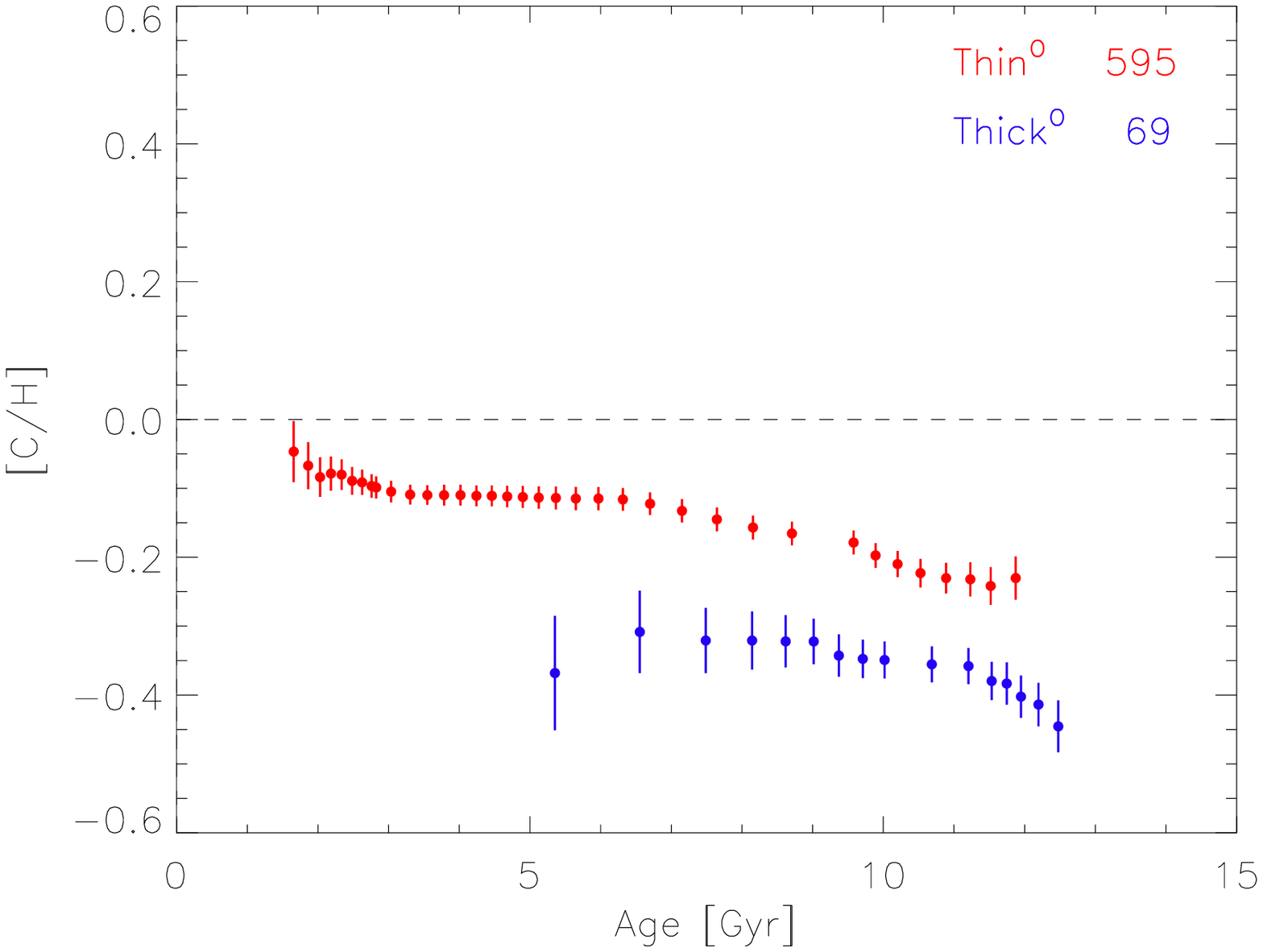} &
\includegraphics[angle=90,width=0.30\textwidth]{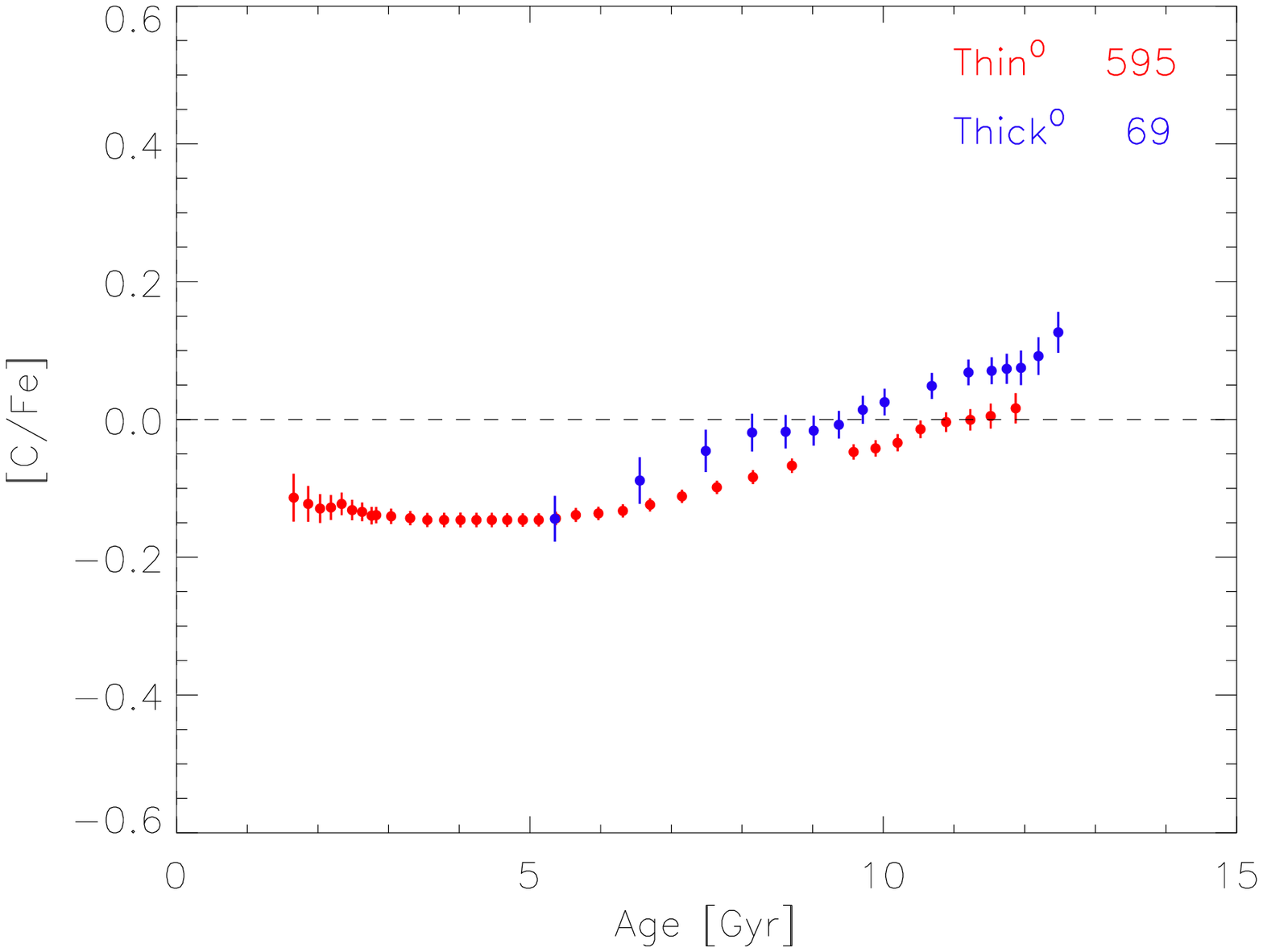} &
\includegraphics[angle=90,width=0.30\textwidth]{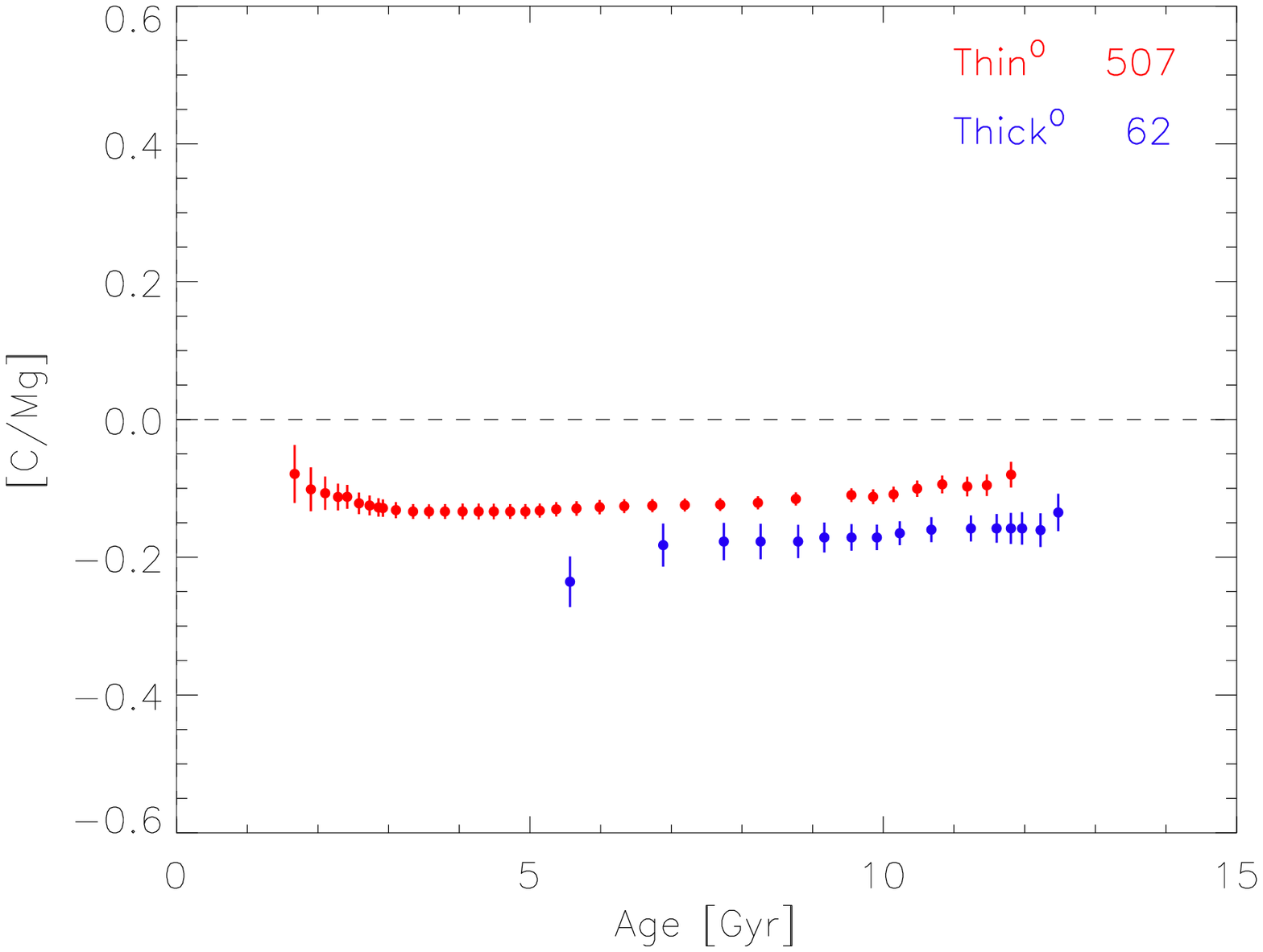}\\
\includegraphics[angle=90,width=0.30\textwidth]{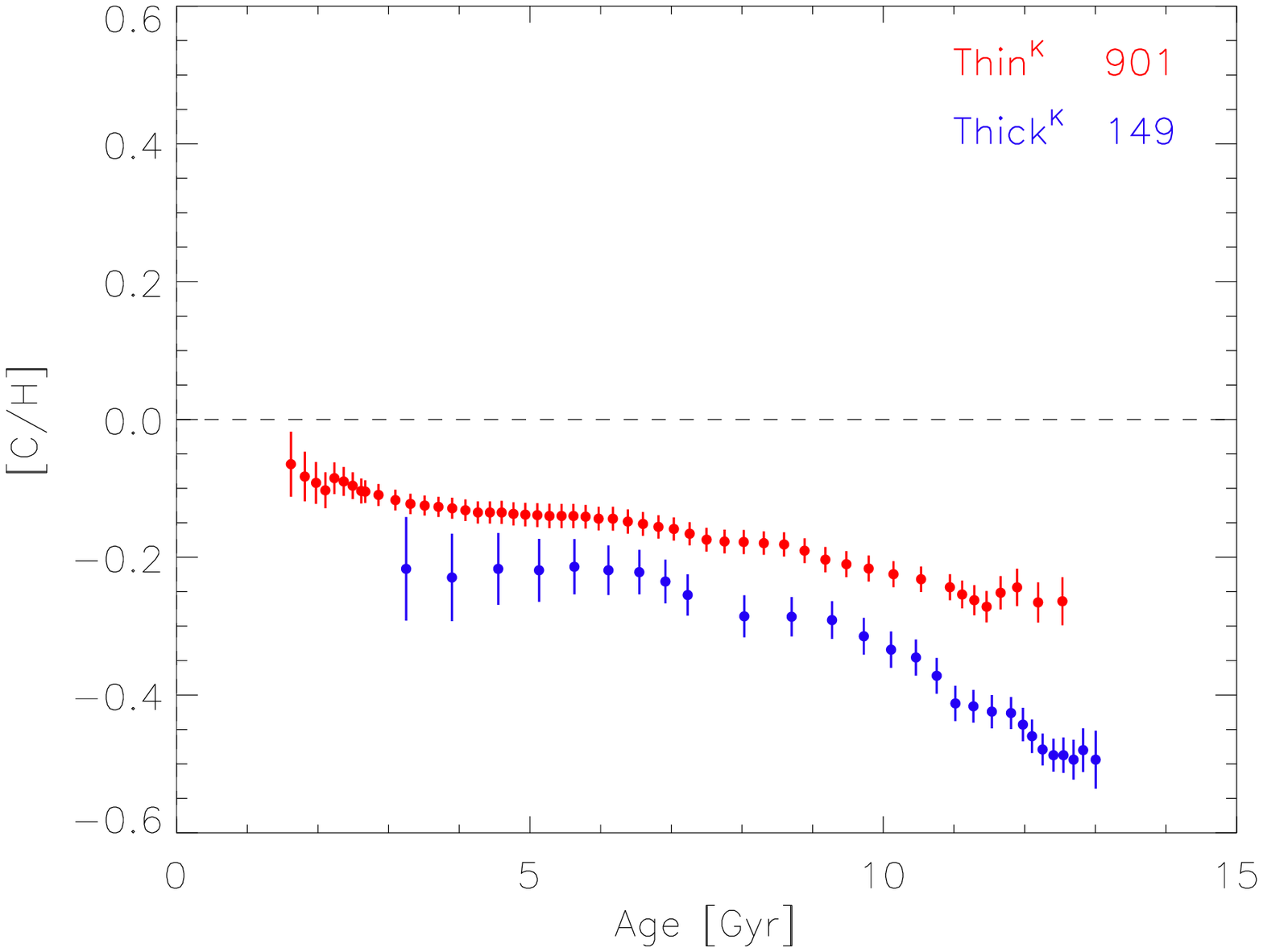} &
\includegraphics[angle=90,width=0.30\textwidth]{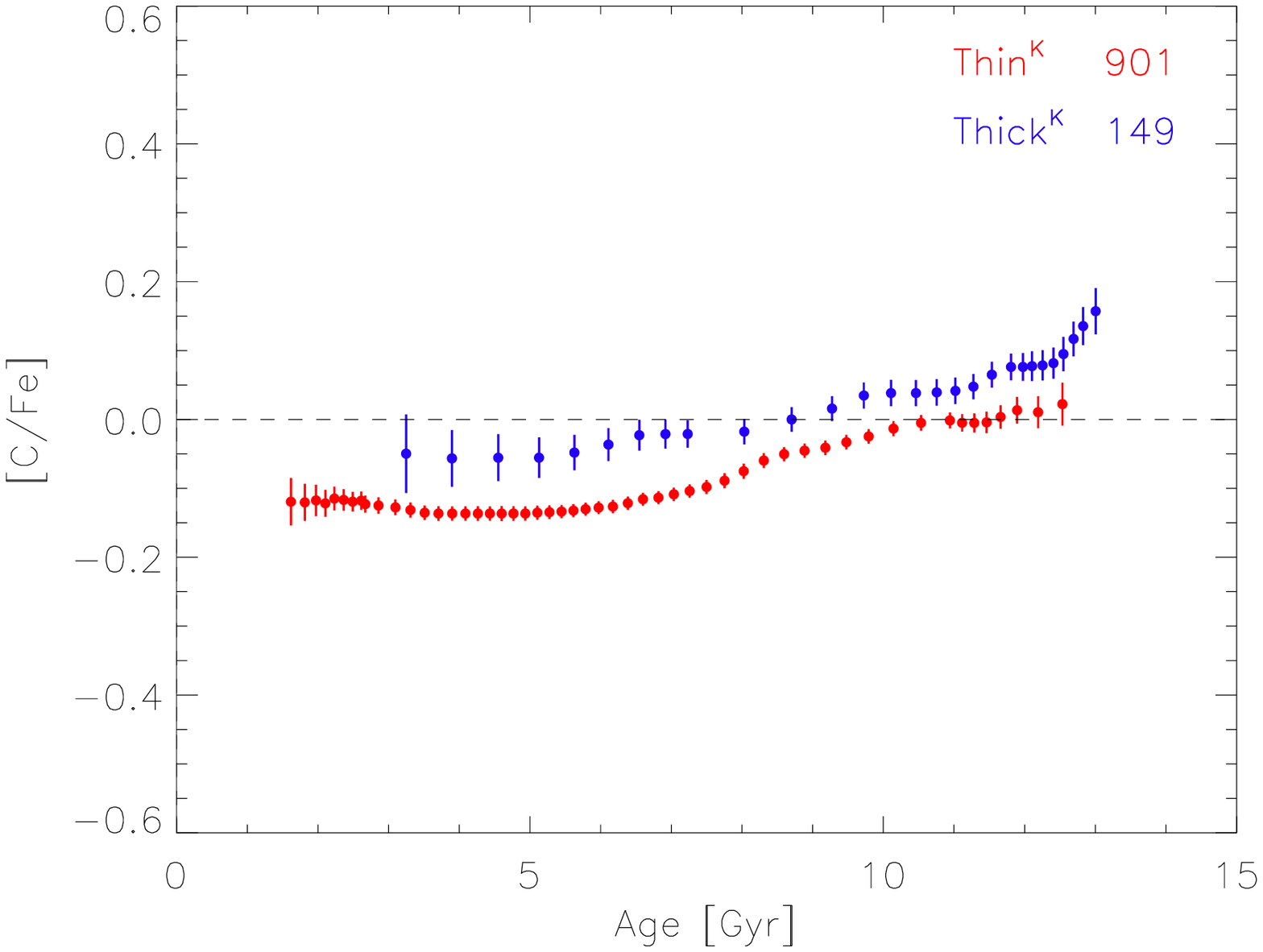} &
\includegraphics[angle=90,width=0.30\textwidth]{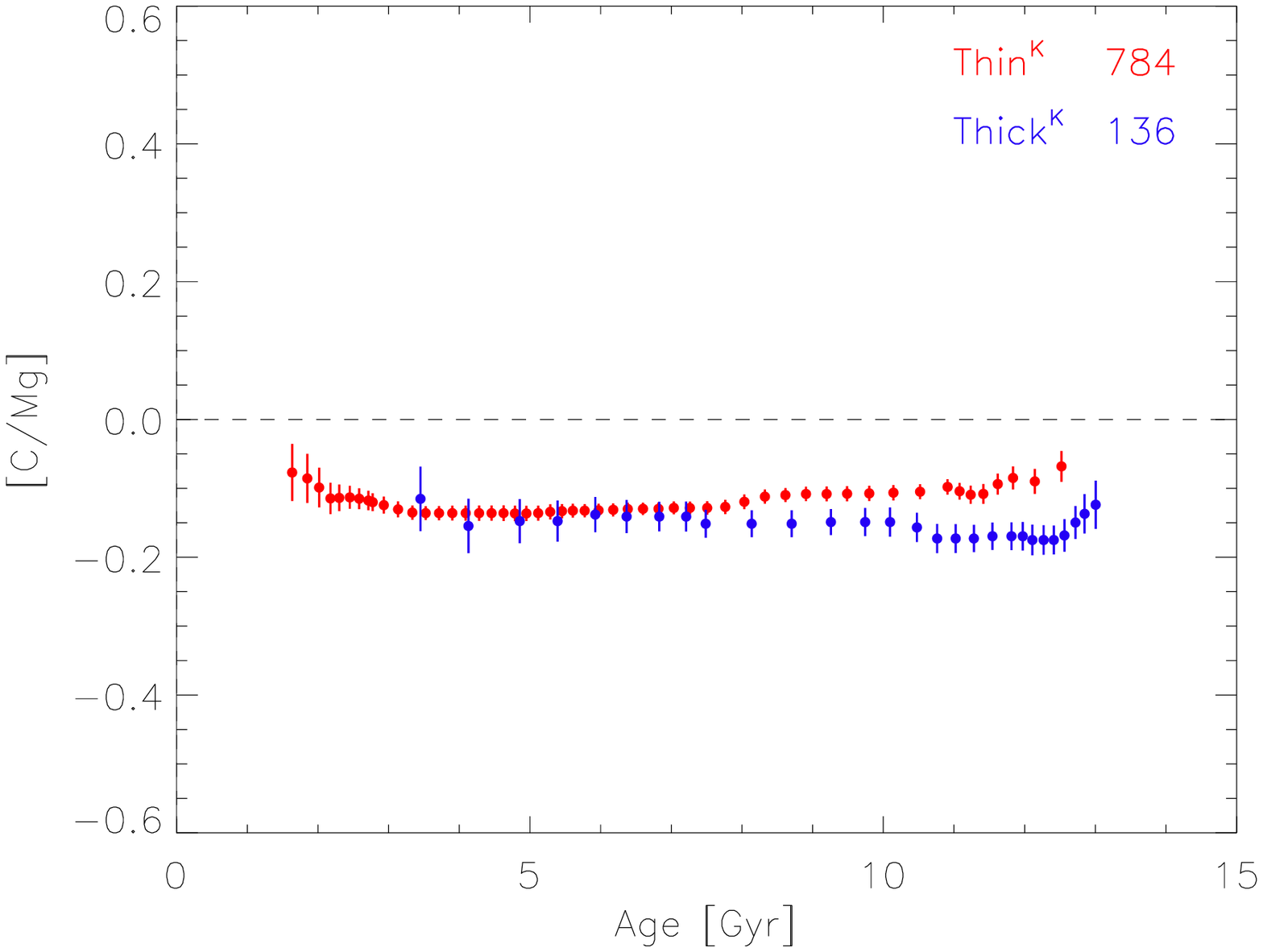}\\
\includegraphics[angle=90,width=0.30\textwidth]{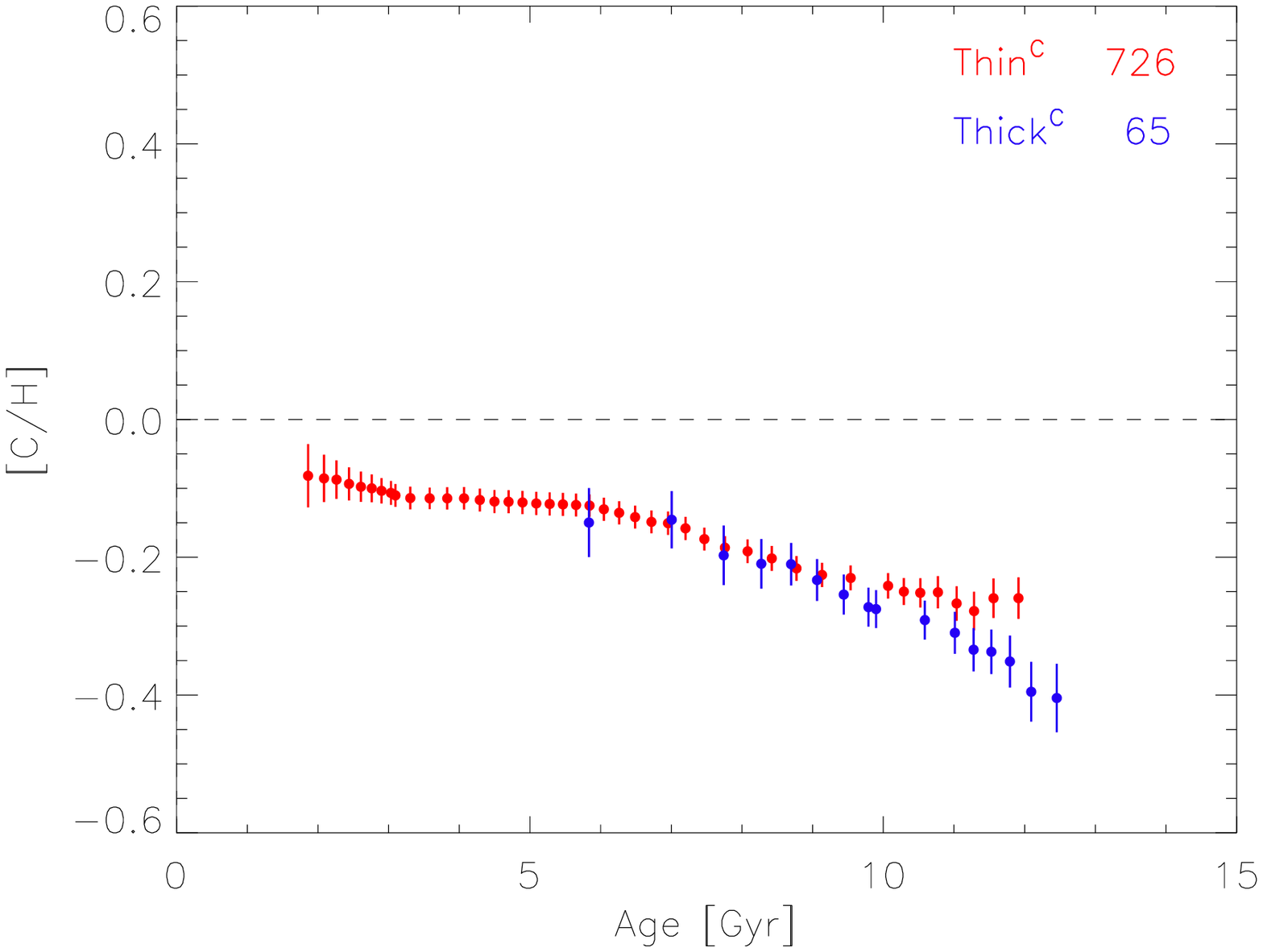} &
\includegraphics[angle=90,width=0.30\textwidth]{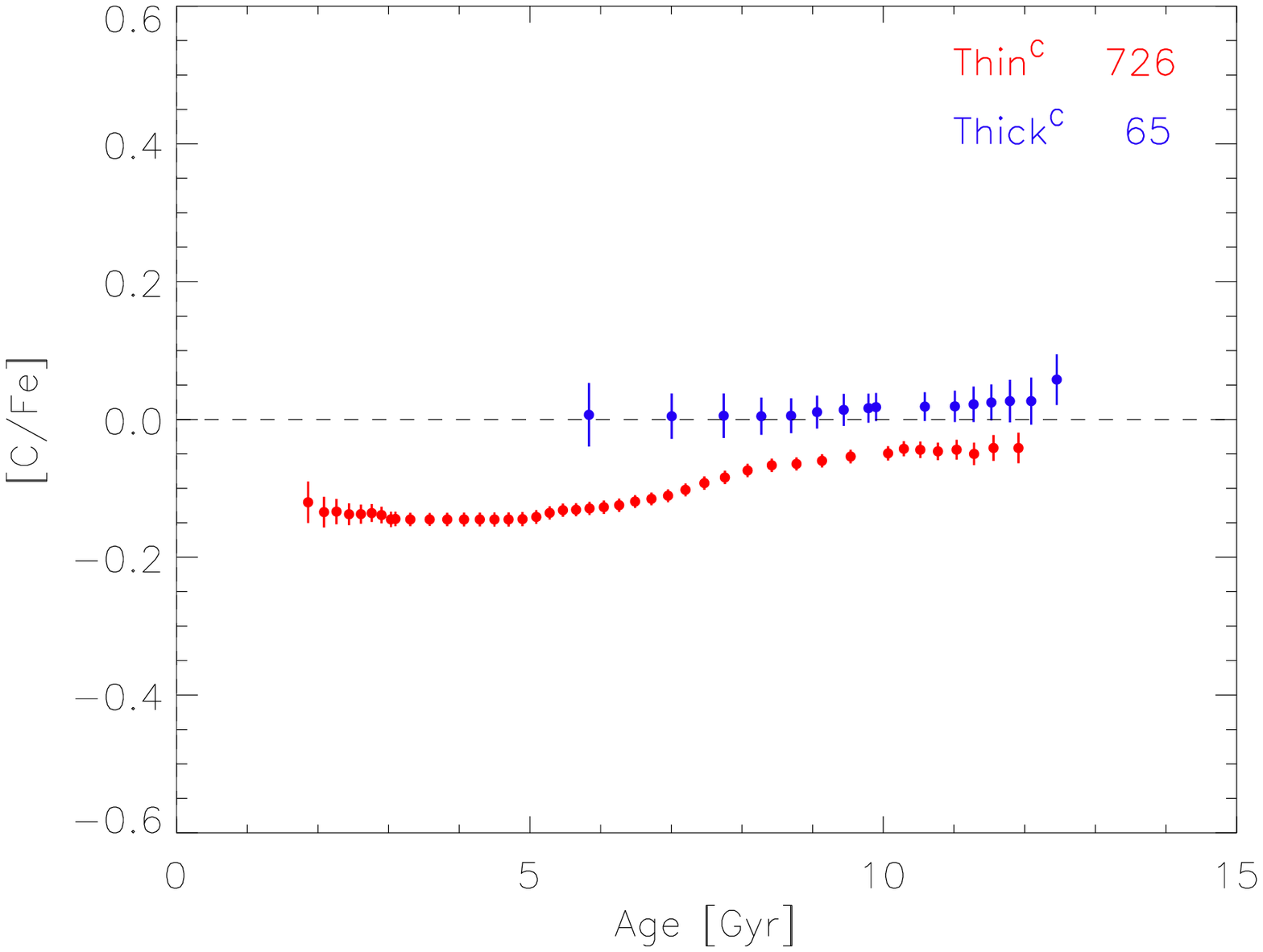} &
\includegraphics[angle=90,width=0.30\textwidth]{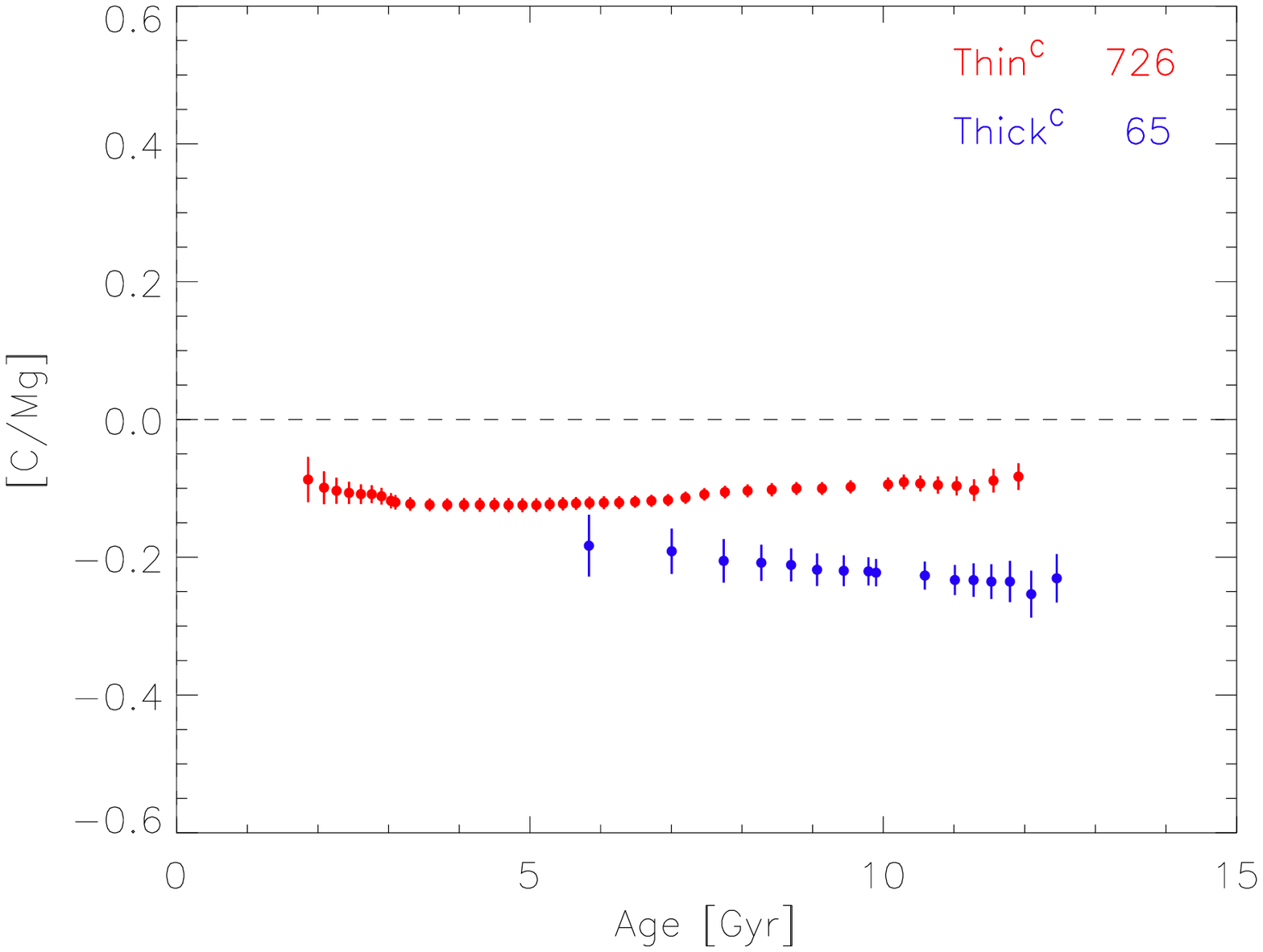}\\
\end{tabular}
\caption{[C/H]  (upper panels),  [C/Fe] (middle panels), and [C/Mg] (lower panels) vs Age diagrams for  thin (red) and thick (blue) samples.
Left panels:   Thin$^{\rm C}$ and  Thick$^{\rm C}$ samples;  central panels:  Thin$^{\rm K}$ and  Thick$^{\rm K}$ samples; right panels: Thin$^{\rm O}$ and  Thick$^{\rm O}$ samples. 
\label{fig:Age_Abu}}
\end{center}
\end{figure}

\begin{figure}[htbp]
\includegraphics[width=0.5\textwidth]{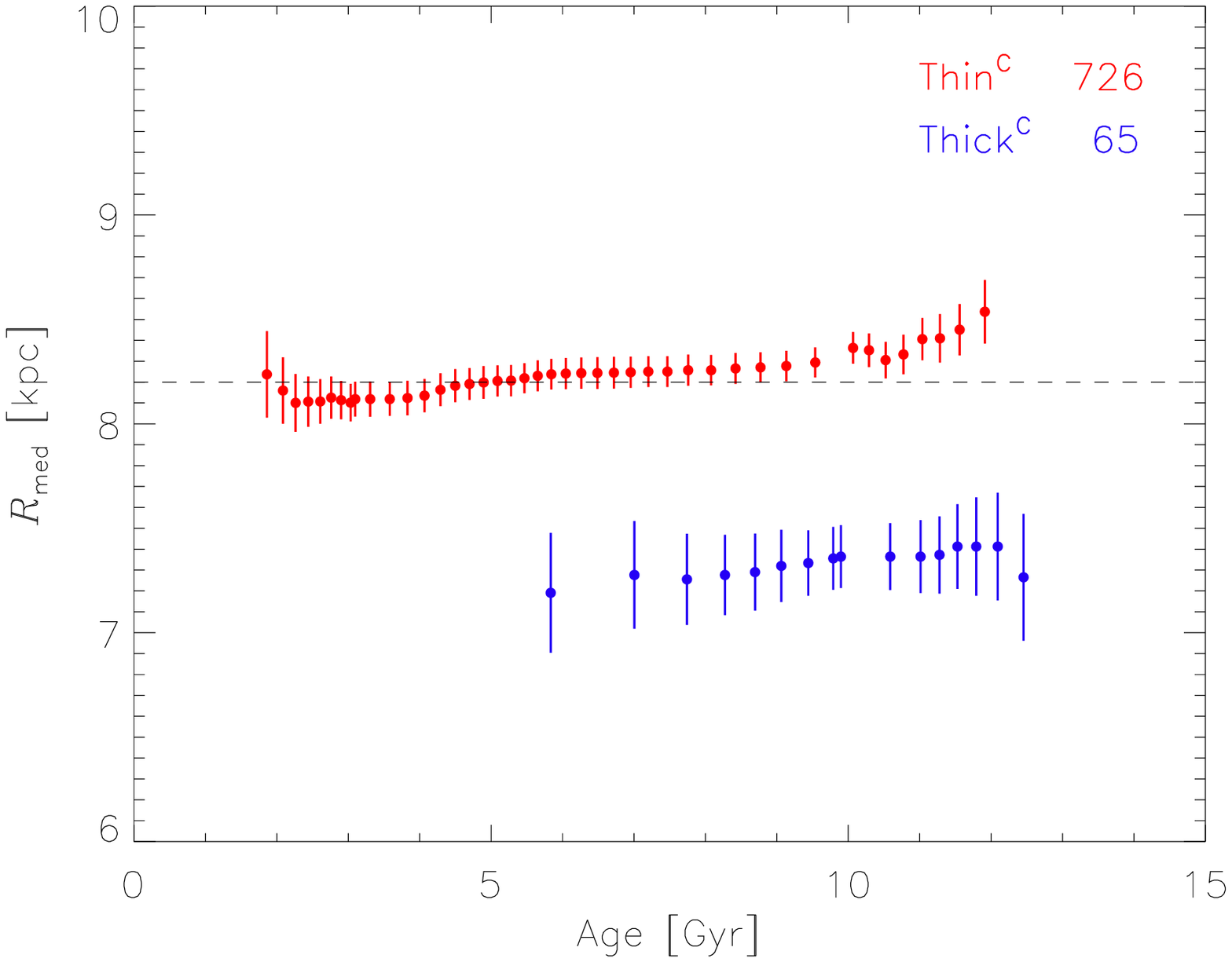}
\includegraphics[width=0.5\textwidth]{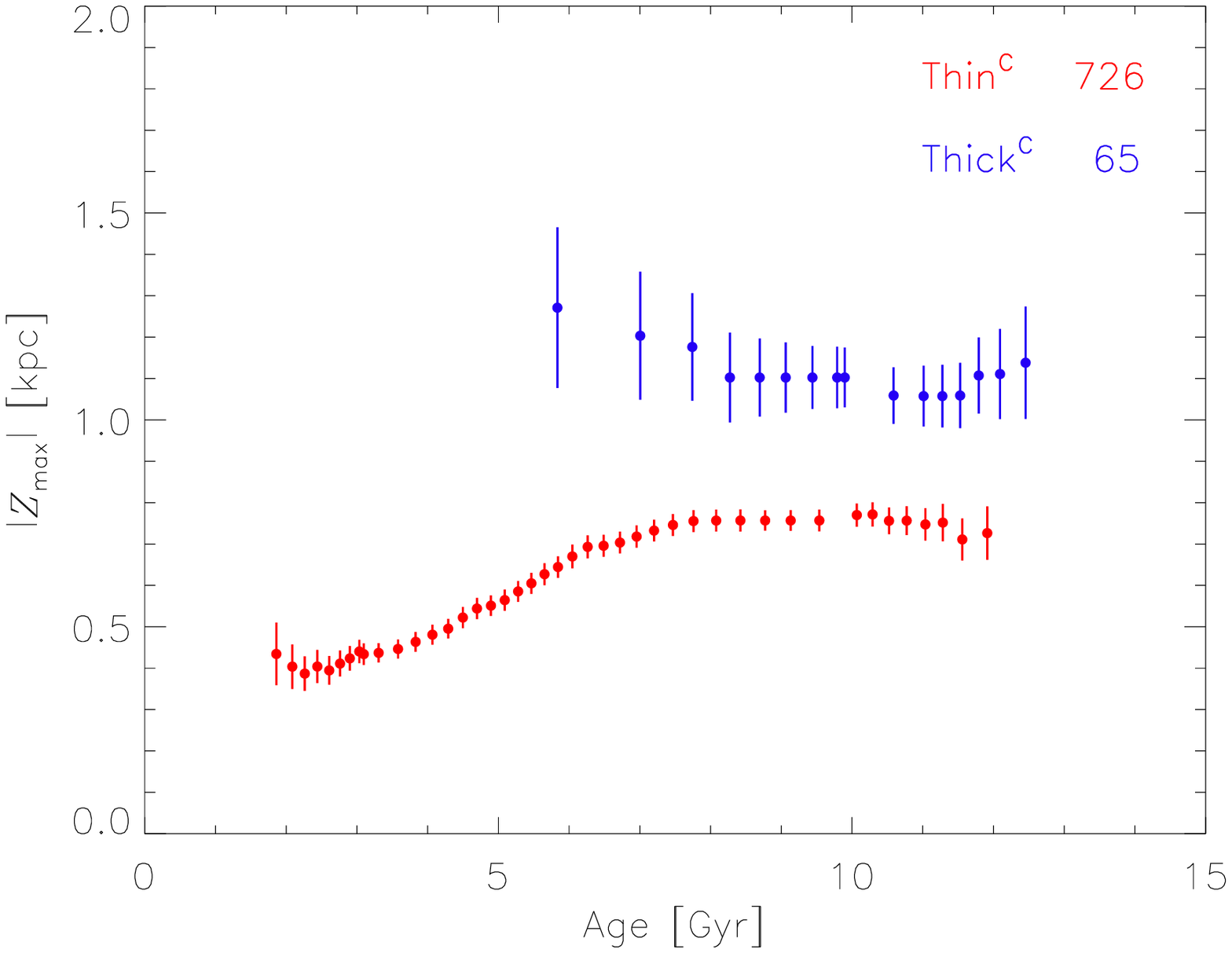}
\includegraphics[width=0.5\textwidth]{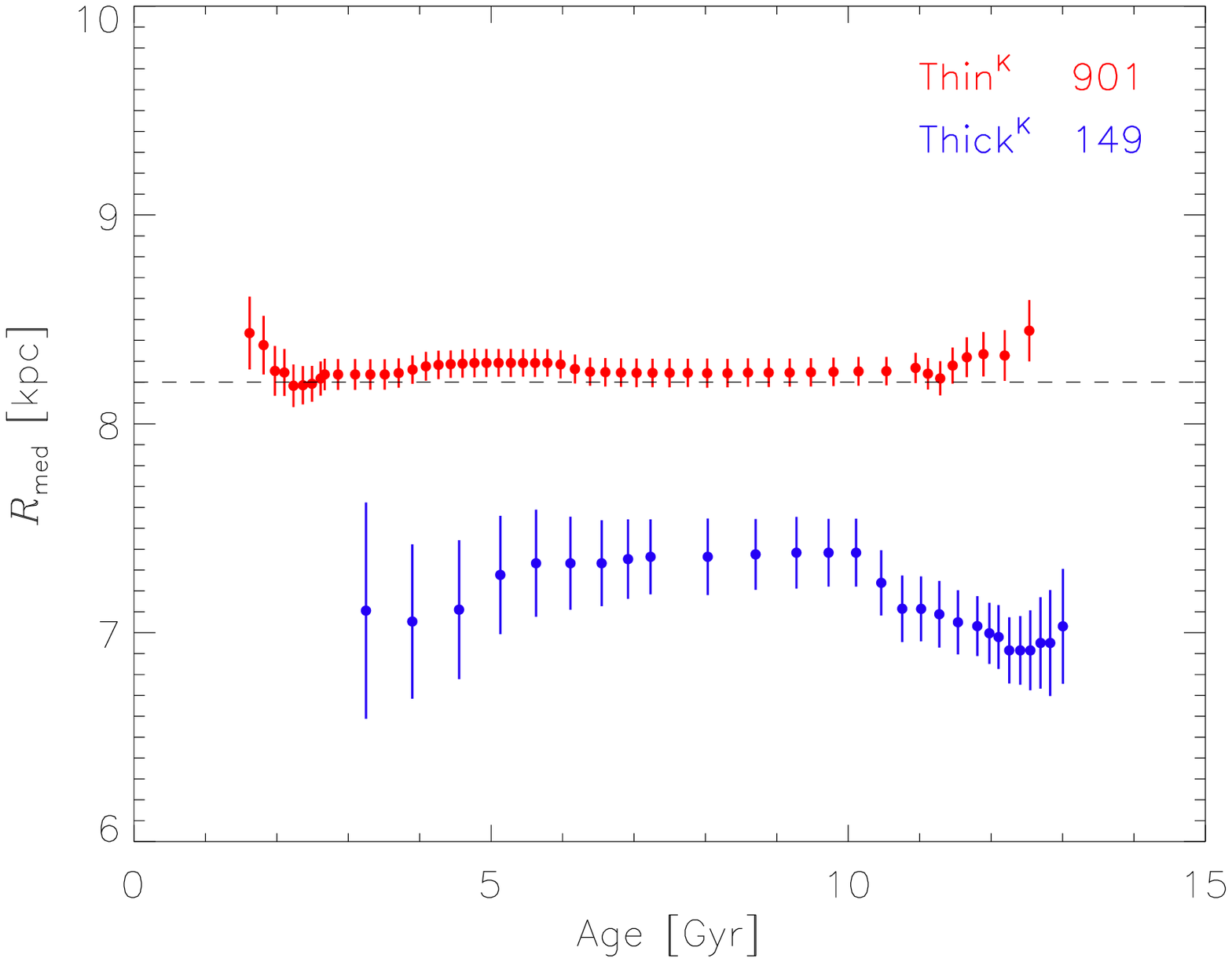}
\includegraphics[width=0.5\textwidth]{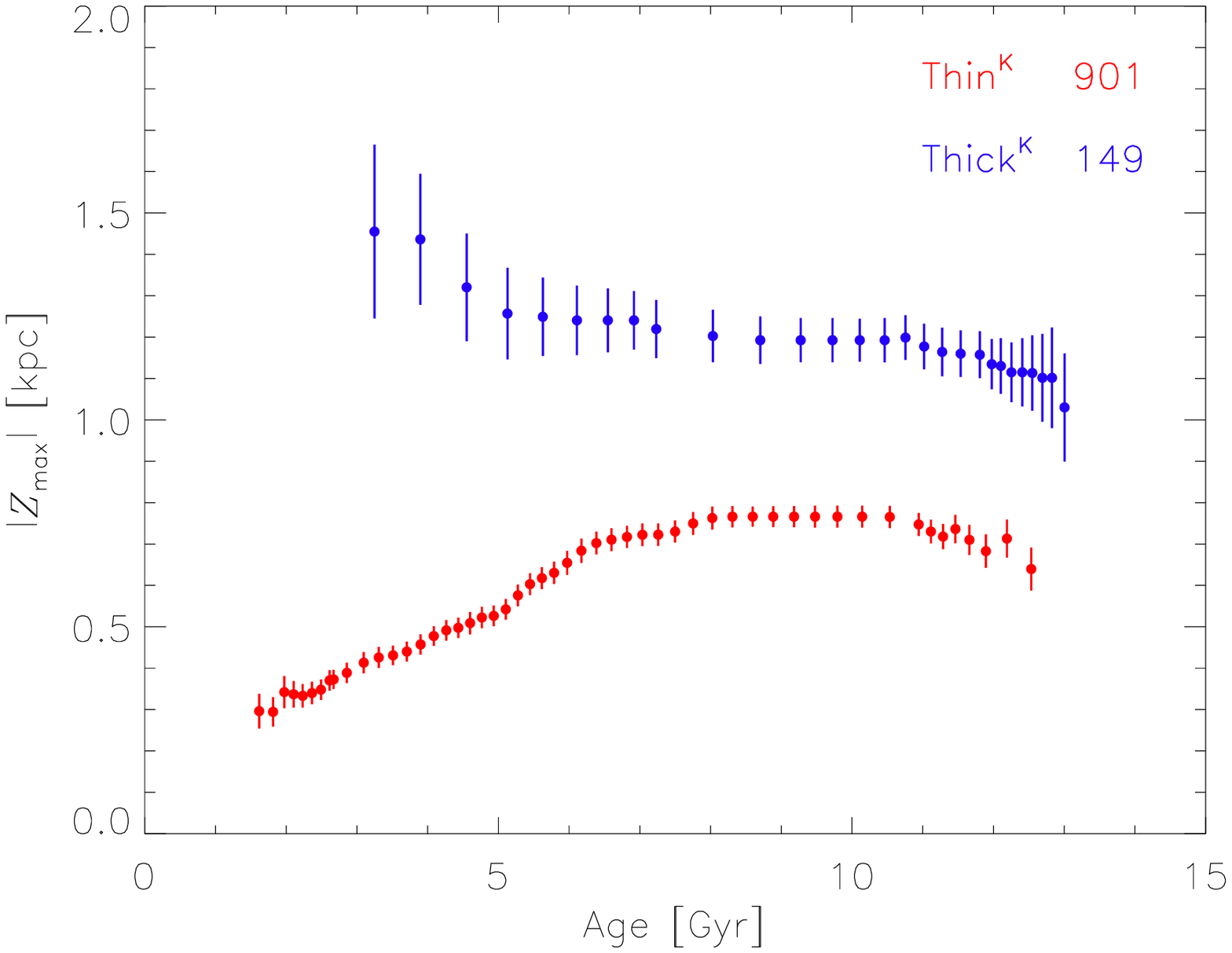}
\includegraphics[width=0.5\textwidth]{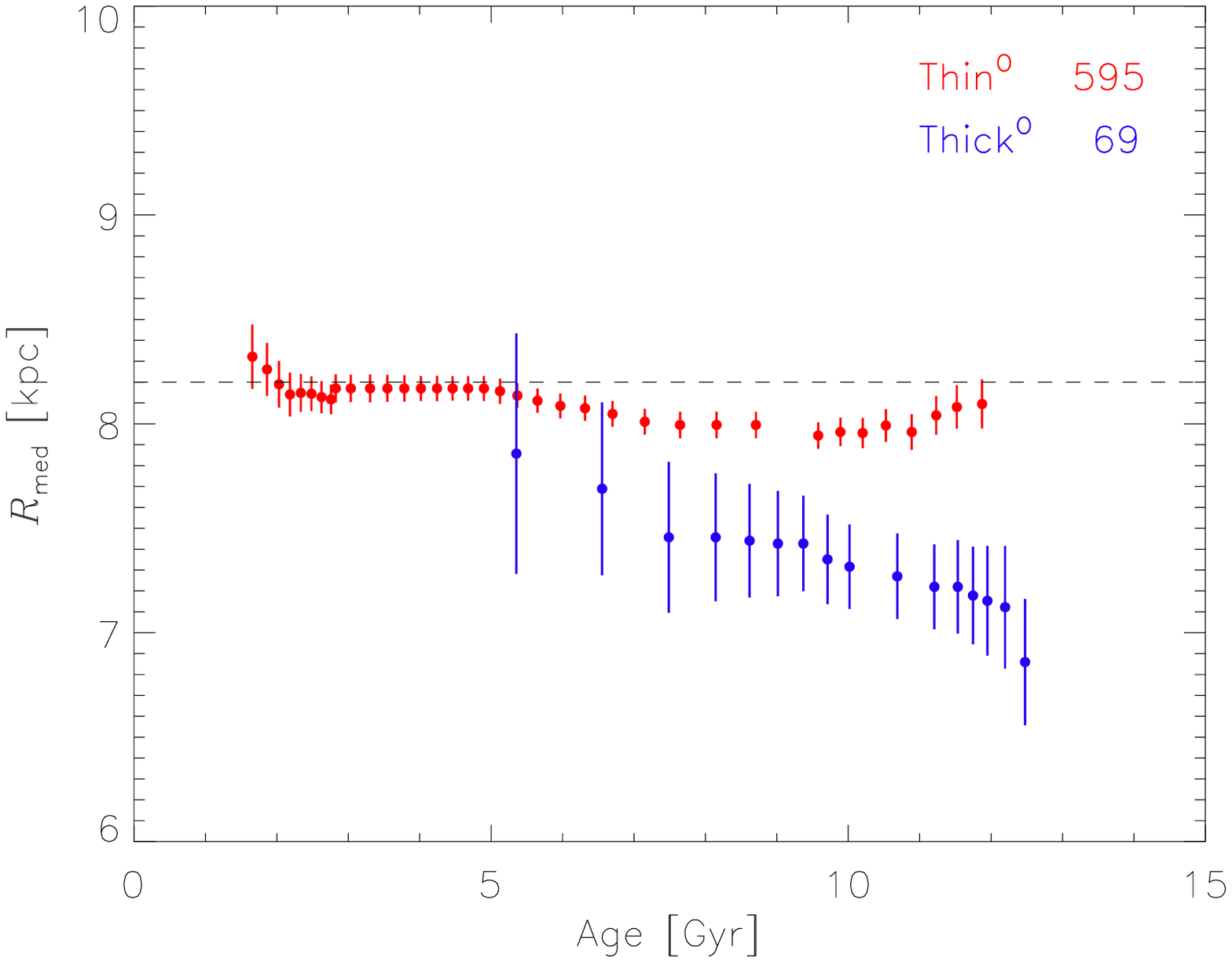}
\includegraphics[width=0.5\textwidth]{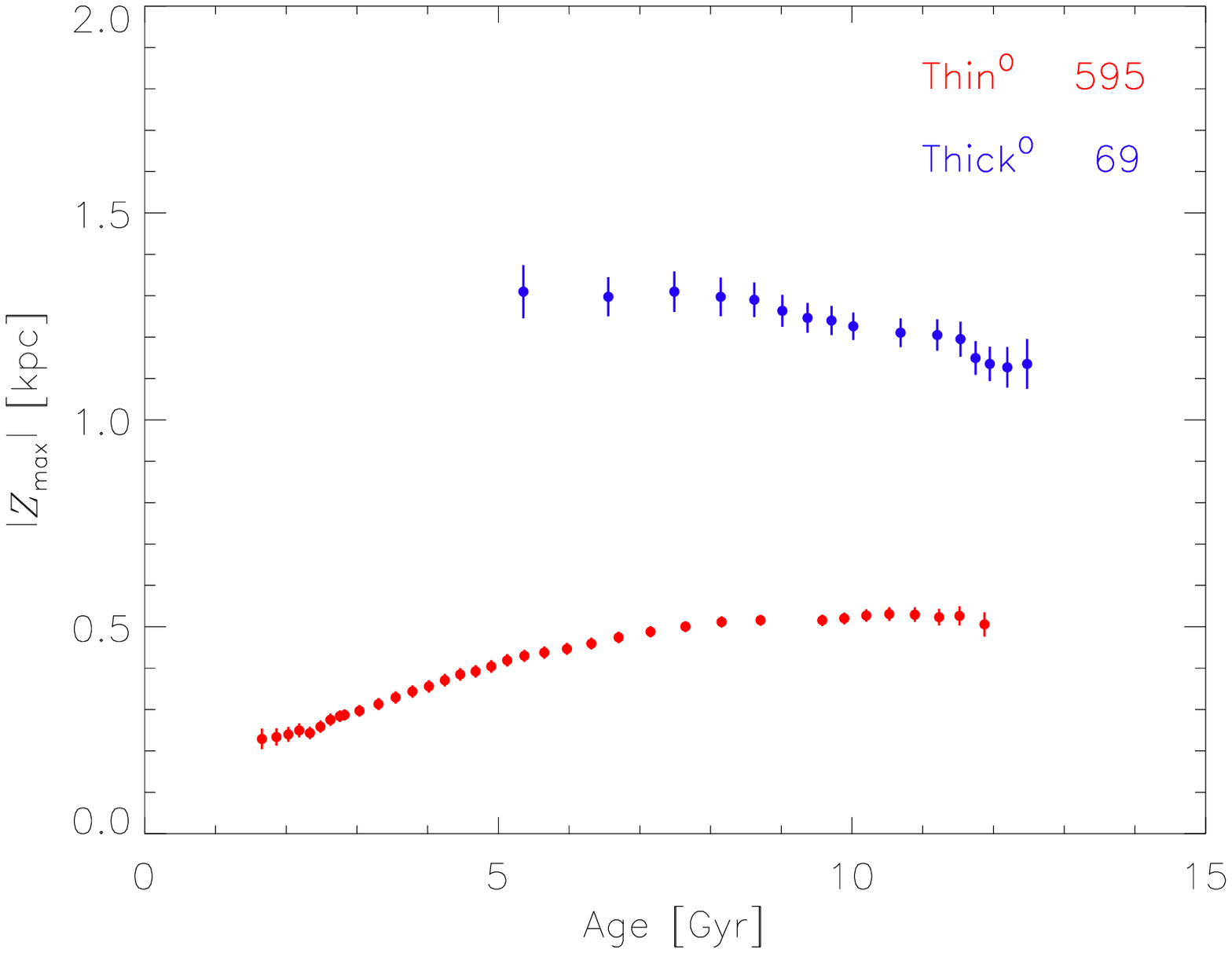}
\caption{$R_{\rm med}$  (left panels) and $|Z_{\rm max}|$  vs Age diagrams  for  thin (red) and thick (blue) samples. 
Upper panels:   Thin$^{\rm C}$ and  Thick$^{\rm C}$ samples;  middle panels:  Thin$^{\rm K}$ and  Thick$^{\rm K}$ samples; lower panels:  Thin$^{\rm O}$ and  Thick$^{\rm O}$ samples.
\label{fig:Age_Rmed}}
\end{figure}

Figures\,\ref{fig:Age_Abu} and \ref{fig:Age_Rmed} show the trends of [C/H], [C/Fe], [C/Mg], $R_{\rm med}$, and $|Z_{\rm max}|$ versus age for the thin (red) and thick (blue) disk samples selected chemically, kinematically, and using orbit characteristics. 

\begin{figure}[htbp]
\includegraphics[width=0.5\textwidth]{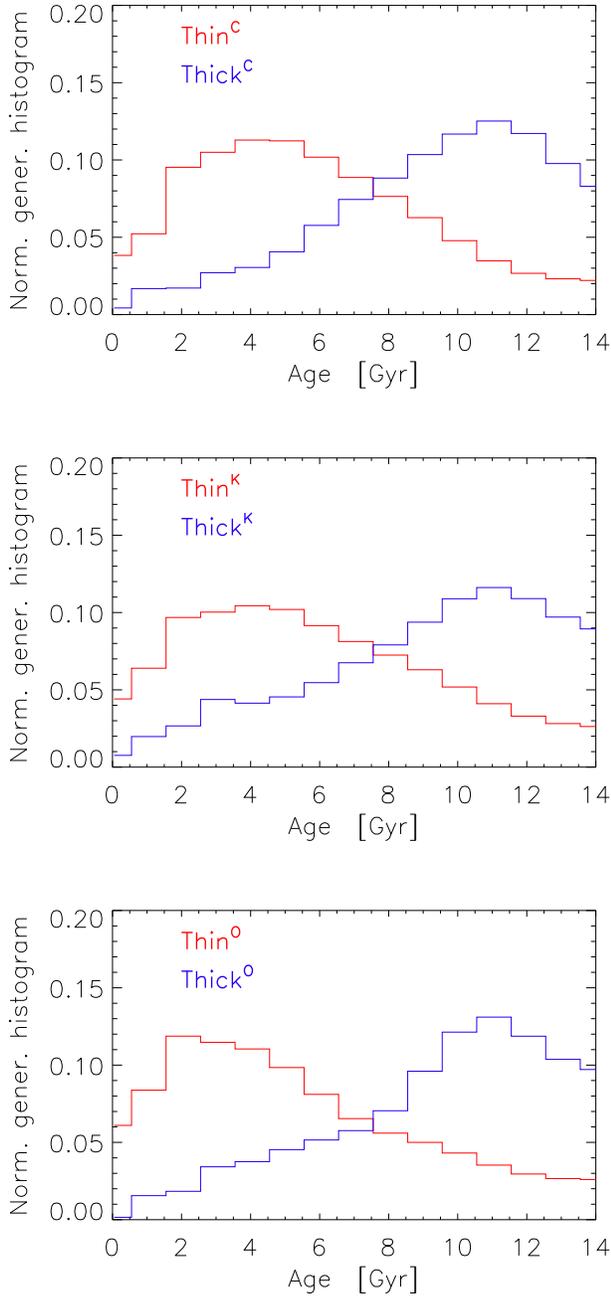}
\caption{Normalized generalized Age histograms   for  thin (red) and thick (blue) disk samples. 
Upper panels:   Thin$^{\rm C}$ and  Thick$^{\rm C}$ samples;  middle panels:  Thin$^{\rm K}$ and  Thick$^{\rm K}$ samples; lower panels:  Thin$^{\rm O}$ and  Thick$^{\rm O}$ samples.
\label{fig:histo_Age}}
\end{figure}

The content of the Figures\,\ref{fig:Age_Abu} and \ref{fig:Age_Rmed} can be summarised as follows:
\begin{itemize}
    \item all the panels show, for any of the three selection i.e. for each pair of samples,  that the thick disk stars are, on the average, older than the thin disk ones (see also Figure\,\ref{fig:histo_Age} where the extended wings of the normalized generalized age histograms, built by summing the individual age probability distributions, at low ages for the thick disk stars and at high ages for the thin disk stars are probably spurious features due to the large uncertainties affecting individual stellar ages). This is in agreement with the common understanding of Galactic chemical evolution based on serial and two-infall models \citep[][and references therein]{GRI17} which predict that the thick disk formed before the thin one. Also in a cosmological context the formation of the thick and thin disk can be explained by means of  an early and later accretion of gas, respectively \citep[][and references therein]{CAL09, SPI19}, with a delay between two episodes which typically is of a few Gyr.
    \item according to our results thin disk stars span an age range from $\sim2$ to  $\sim12$\,Gyr, which would indicate that the formation of the thin disk took place about 2--3\,Gyr after the initial stages of the Milky Way evolution. This is in qualitative agreement with chemical abundance studies matched with asteroseismologic age determinations, which indicate a delay of $\sim 4$\,Gyr between the first and second accretion episodes which gave place to the MW disk \citep{SPI19}.
    \item the thick disk stars span an age range from $\sim5$ to  $\sim13$\,Gyr. Therefore, our results show hints that the thick disk started forming about 2\,Gyr before the thin disk and that its formation lasted about 6-8\,Gyr; 
    \item the oldest thick disk stars have lower [C/H] than the thin disk stars. This is more evident for the selection based on orbital parameters and less evident for the chemical selection. The trend with age shows a steeper increase for both the oldest and the youngest stars suggesting that C is produced at the beginning  by  massive stars and in more recent time by low mass stars or by high metallicity massive stars due to their enhanced mass loss. This is in agreement with the results of \citet{NIS14} and with the predictions of the ``best fitting model'' by \citet{CAR05}; 
    \item thick disk stars have higher [C/Fe] than thin disk stars for all the selections. The thick disk trends show a monotonic  decrease of [C/Fe] with decreasing age. On the other hand, the thin disk trends show a decrease with age from 10\,Gyr to about 5\,Gyr, a flattening from 5\,Gyr to 3\,Gyr, and then an hint of uprising for the youngest stars suggesting again that there is an extra source of C at more recent time due to low mass stars;
    \item both thin and thick disk stars show almost flat trends of [C/Mg] with age and an increase of the ratio for the youngest thin disk stars. The average [C/Mg] difference between the two chemically selected samples is, to some extent, expected because of the choice of identifying thin and thick disk stars with high and low [Mg/Fe] stars, respectively. 
    However, the increase of [C/Mg] for the youngest stars is a plausible evidence that low mass stars or massive stars at high metallicity due to enhanced mass loss contribute a significant amount of C at recent times;
    \item thick disk stars have, on the average, lower $R_{\rm med}$ and higher $|Z_{\rm max}|$ than thin disk stars. 
    \begin{itemize}

    \item the thin disk stars span an $R_{\rm med}$ range from $\sim8.0$ to  $\sim8.5$\,kpc with an almost flat behaviour; 
    \item the thick disk stars span an $R_{\rm med}$ range from $\sim7$ to  $\sim7.5$\,kpc showing that the oldest stars were formed at smaller $R_{\rm med}$. These are the diagrams ($R_{\rm med}$ vs age) which show the largest differences between the differently selected samples in particular for the thick disk stars and are also the less reliable ones since stellar migration can prevent us to use our computed $R_{\rm med}$ values as indicators of the stellar birth places; 
    \item the thin disk stars span  a $|Z_{\rm max}|$ range from $\sim0.3$ to  $\sim0.8$\,kpc (but for the sample selected using orbital parameters which is confined below 0.6) with a trend of decreasing $|Z_{\rm max}|$ with decreasing age starting at $\sim7$\,Gyr;
    \item the thick disk stars span a $|Z_{\rm max}|$ range from $\sim1.1$ to  $\sim1.5$\,kpc.  The trend with age is almost flat with some hints of increasing $|Z_{\rm max}|$ for decreasing age. The larger separation in $|Z_{\rm max}|$ between the thin and the thick disk stars for the selection based on orbital parameters is an effect of the selection itself and, therefore, can be artificial.
  \end{itemize}
These results are in agreement with \citet{BEN14} who did not find any low-$\alpha$ star at $R_{\rm med}<7$\,kpc and with \citet{KOR15} who found very few low--$\alpha$ stars for $R<7.5$\,kpc. Assuming that $R_{\rm med}$ is a measure of the distance from the Galactic centre of the stellar birthplace, its increase with decreasing age for thick disk stars can be explained, as an evidence for ``inside--out'' \citep{NUZ19} formation scenarios as found also by \citet{BER14}. This fact, together with the $|Z_{\rm max}|$ versus age behaviours  that suggests an ``upside--down'' \citep{FRE17} formation scenario of the disk components of our Galaxy, indicates that, radially,  the central disk was formed before the outer disk, and, vertically, the thick disk was formed before the thin disk. 

\end{itemize}

\begin{deluxetable}{c|rrrrr|rrrrr}[htbp]
\tablecaption{Percentages of stars in common among the different selection samples. \label{tab:percentage}}
\tablecolumns{11}
\tablewidth{0pt}
\tablehead{
\multicolumn{1}{c|}{} & \multicolumn{5}{c|}{Thin} &\multicolumn{5}{c}{Thick} \\
\hline
\multicolumn{1}{c|}{Selection} &
\colhead{N$_{\rm star}$} &
\colhead{C} &
\colhead{K} &
\colhead{O} &
\multicolumn{1}{c|}{All} &
\colhead{N$_{\rm star}$} &
\colhead{C} &
\colhead{K} &
\colhead{O} &
\colhead{All} \\[-8pt]
\multicolumn{1}{c|}{} &
\colhead{} &
\colhead{\%} &
\colhead{\%} &
\colhead{\%} &
\multicolumn{1}{c|}{\%} &
\colhead{} &
\colhead{\%} &
\colhead{\%} &
\colhead{\%} &
\colhead{\%}
}
\startdata
C & 1267 &    &  72 & 50 & 48  &  99 &    & 32  & 14  & 12  \\
K & 1356 & 67 &     & 65 & 45  & 196 & 16 &     & 36  &  6 \\
O & 934  & 68 &  95 &    & 66  &  90 & 15 & 80  &     & 13 \\
\enddata
\tablecomments{~~C--chemical selection; K--kinematical selection; O--orbital selection.}
\end{deluxetable}

\section{Conclusions}
\label{sec:conclusion}
In this paper we investigated the carbon abundance in the thin and thick disk of our Galaxy. The analysis is based on a sample of 2133 dwarf stars from the Gaia-ESO survey. Their carbon abundances were derived by comparing the observed UVES spectra with ``on-the-fly'' computed synthetic spectra obtained from fully consistent atmosphere models. 
The designation of stars to the thin or thick disk populations was addressed by adopting three different selection approaches, i.e. a chemical one based on positions in the [Mg/Fe]-[Fe/H] plane, a kinematical one based on stellar Galactic velocities, and a third one based on orbital parameters.

The three different selection led to different samples of candidate thin and thick disk stars:
\begin{itemize}
    \item chemical selection identified 1267 thin disk stars (Thin$^{\rm C}$ sample) and   99 thick disk stars  (Thick$^{\rm C}$ sample);
    \item kinematical selection identified 1356 thin disk stars (Thin$^{\rm K}$ sample) and   196 thick disk stars  (Thick$^{\rm K}$ sample);
   \item selection based on orbital parameters identified 934 thin disk stars (Thin$^{\rm O}$ sample) and   90 thick disk stars  (Thick$^{\rm O}$ sample).
\end{itemize}
Only 620 and 12 stars  are classified as thin or thick disk stars, respectively, by using all the three selections. The low number of thick disk stars with unanimous classification is, in particular, due to the poor agreement between the chemical selection, i.e. high-[Mg/Fe],  and each of the other two (see Table\,\ref{tab:percentage}).
Even if the different selections produced not fully concordant  lists of candidates, the chemical and kinematical general trends of the thin and thick $^{\rm C,K,O}$ samples display quite common behaviours and our results show that:
\begin{itemize}
    \item in all the cases, our thin and thick disk stars show different  carbon abundances [C/H],  [C/Fe], and [C/Mg] abundances ratios;
\item  our thin and thick disk stars span different age intervals, with the latter being, on  average, older than the former; 
\item the behaviours of [C/H], [C/Fe], and [C/Mg] versus [Fe/H], [Mg/H], and age all suggest that C is primarily produced in massive stars like Mg but the rise of [C/Mg] for young thin disk stars indicates that also low-mass stars may play a role in providing carbon in the Galactic thin disk;
\item the analysis of the orbital parameters $R_{\rm med}$ and $|Z_{\rm max}|$ support an ``inside--out'' and ``upside--down'' formation scenario for the disks of Milky Way.
\end{itemize}

The data used in this paper, together with the derived atmospheric parameter values, are parts of the full dat--set from the GES survey and will be published through the ESO archive as required for any ESO Public Surveys. All the GES spectra will be publicly available early in 2020, the astrophysical parameters and abundances shortly thereafter.

\acknowledgments
This work is based on data products from observations made with ESO Telescopes at the La Silla Paranal Observatory under programme ID 188.B-3002. 
These data products have been processed by the Cambridge Astronomy Survey Unit (CASU) at the Institute of Astronomy, 
University of Cambridge, and by the FLAMES/UVES reduction team at INAF/Osservatorio Astrofisico di Arcetri. 
These data have been obtained from the Gaia-ESO Survey Data Archive, prepared and hosted by the Wide Field Astronomy Unit, Institute for Astronomy, 
University of Edinburgh, which is funded by the UK Science and Technology Facilities Council.
This work was partly supported by the European Union FP7 programme through ERC grant number 320360 and 
by the Leverhulme Trust through grant RPG-2012-541. We acknowledge the support from INAF and Ministero dell' Istruzione, 
dell' Universit\`a e della Ricerca (MIUR) in the form of the grant ``Premiale VLT 2012''. The results presented here benefit from 
discussions held during the Gaia-ESO workshops and conferences supported by the ESF (European Science Foundation) through the GREAT Research Network Programme.

This work has made use of data from the European Space Agency (ESA) mission Gaia (\url{https://www.cosmos.esa.int/gaia}), processed by the Gaia
Data Processing and Analysis Consortium (DPAC, \url{https://www.cosmos.esa.int/web/gaia/dpac/consortium}). Funding for the DPAC
has been provided by national institutions, in particular the institutions participating in the Gaia Multilateral Agreement.

This work received partial financial support
from PRIN MIUR 2010--2011 project ``The Chemical and dynamical Evolution of the Milky Way
and Local Group Galaxies'', prot. 2010LY5N2T and  by the National Institute for Astrophysics (INAF) through the grant PRIN-2014 (``The Gaia-ESO Survey''). 
M.C. thanks financial support from CONACyT grant CB-2015-256961. T.B. was supported by the project grant ``The New Milky Way'' from the Knut and Alice Wallenberg Foundation. U.H. acknowledges support from the Swedish National Space Agency (SNSA/Rymdstyrelsen). S.G.S acknowledges the support by Funda\c{c}\~{a}o para a Ci\~{e}ncia e Tecnologia (FCT) through national funds and a research grant (project ref. UID/FIS/04434/2013, and PTDC/FIS-AST/7073/2014). S.G.S. also acknowledge the support from FCT through Investigador FCT contract of reference IF/00028/2014 and POPH/FSE (EC) by FEDER funding through the program ``Programa Operacional de Factores de Competitividade -- COMPET''. 

This research uses the facilities of the Italian Center for Astronomical Archive (IA2) operated by INAF.

%

\vspace{5mm}
\facilities{VLT:Kueyen, UVES}


\software{SPECTRUM  \citep[v2.76f;][]{GRA94}, ATLAS12 \citep{KU05a}
          }

\bibliographystyle{aasjournal}
\bibliography{mybiblio}

\end{document}